\documentclass[%
 reprint,
 superscriptaddress,
 nobibnotes,
 amsmath,amssymb,
 prappl,
]{revtex4-2}

\usepackage{hyperref}
\usepackage{graphicx}
\usepackage{dcolumn}
\usepackage{bm}
\usepackage{multirow}
\usepackage{booktabs}
\usepackage{rotfloat}
\usepackage{esint}

\usepackage{graphicx}
\usepackage{psfrag,color}
\usepackage{lipsum}

\usepackage{xcolor}
\usepackage[export]{adjustbox}

\usepackage{scalerel}

\newcounter{paticounter}

\makeatother

\begin{document}
\preprint{APS/123-QED}

\title{Space-Time Fresnel Prism}

\author{Zhiyu Li}
\email{lizhiyu@stu.xjtu.edu.cn}%
\affiliation{
 State Key Laboratory of Electrical Insulation and Power Equipment, Department of Electrical Engineering, Xi’an Jiaotong University, Xi’an, Shaanxi, China
}%
\affiliation{
 Department of Electrical Engineering, KU Leuven, Leuven, Belgium
}%
 
\author{Xikui Ma}
\affiliation{
 State Key Laboratory of Electrical Insulation and Power Equipment, Department of Electrical Engineering, Xi’an Jiaotong University, Xi’an, Shaanxi, China
}%
 
\author{Amir Bahrami}%
\affiliation{
  Department of Electrical Engineering, KU Leuven, Leuven, Belgium
}%

\author{Zo{\'e}-Lise Deck-L{\'e}ger}%
\affiliation{
  Department of Electrical Engineering, Polytechnique Montr{\'e}al, Montr{\'e}al, Quebec, Canada
}%

\author{Christophe Caloz}
\affiliation{
  Department of Electrical Engineering, KU Leuven, Leuven, Belgium
}%

\date{\today}

\begin{abstract}
Space-time modulation-based metamaterials have recently spurred considerable interest, owing to the fundamental addition of the time dimension to the medium parameters, and resulting novel properties and potential applications. However, the implementation of most related structures -- e.g., involving step, slab or gradient discontinuities -- has been hindered by the impossible requirement of infinitely or prohibitively large device sizes. We provide here a solution to this issue, consisting in a space-time transposition of the conventional Fresnel prism, whereby a copy of the target modulation is periodically re-injected at the input of a Fresnel-reduced finite structure, so as to provide the same anharmonic and nonreciprocal frequency conversion as the target space-time interface in the case of a modulation step. This concept, which may readily extend to slab or gradient modulations, as well as accelerated profiles for space-time chirping operations, may pave the way for the practical development of a wide range of novel microwave and optical space-time systems.
\end{abstract}

\maketitle

\section{Introduction}
Space-time systems are structures that are formed by the modulation of some parameter (e.g., the refractive index) of a host medium in space and time or only in time~\cite{Lurie_2007_mathematical,Caloz_2019_spacetime1,Caloz_2019_spacetime2,Engheta_2021_metamaterials,Caloz_GSTEMs,Engheta_2023_fourdimensional,Alu_2023_scattaringtemporal}. They represent fundamental extensions of pure-space systems and have recently led to a wide diversity of novel applications, including magnetless nonreciprocity~\cite{Yu_2009_opticalisolation,Estep_2014_nonreciprocity,Taravati_2017_nonreciprocal,Li_2018_floquet}, time reversal and refocusing~\cite{Bacot_2016_timereversal,Deck_2018_wave}, inverse-prism scattering~\cite{Akbarzadeh_2018_inverse}, temporal coating~\cite{Pena_2020_tempcoating}, fundamental bound breaking~\cite{Shlivinski_2018_BF,Li_2019_beyondChu}, dynamic crystals~\cite{Deck_2019_uniform,Peng_2022_topological} and light deflection or bending~\cite{Huidobro_2019_fresnel,Bahrami_2023_electrodynamics}.

The modulation may be periodic~\cite{Tien_1958_parametric,Cassedy_1963_dispersion1,Cassedy_1967_dispersion2,Reed_2003_color,Yu_2009_opticalisolation,Sanchez_2009_tempslab,Estep_2014_nonreciprocity,Martinez_2016_tempcrystal,Taravati_2017_nonreciprocal,Li_2018_floquet,Mirmoosa_2019_timereactive,Li_2019_beyondChu,Huidobro_2019_fresnel,Galiffi_2019_broadband,Deck_2019_uniform,Peng_2022_topological,Pendry_2022_photon,Xu_2022_tempGTMM,Bahrami_2023_electrodynamics} or aperiodic~\cite{Morgenthaler_1958_velocity,Felsen_1970_wave,Fante_1971_transmission,Kunz_1980_plane,Mendoncca_2003_temporalsplitter,Biancalana_2007_dynamics,Bacot_2016_timereversal,Akbarzadeh_2018_inverse,Shlivinski_2018_BF,Deck_2018_wave,Pena_2020_tempcoating,Silbiger_2023_filter,Li_2023_TIR}, depending on the application. While periodic-modulation space-time systems may be implemented in finite-size structures, using Bloch-Floquet impedance (reflection-less) terminations, aperiodic-modulation (e.g., step, slab, gradient) space-time systems are practically restricted by a \emph{fundamental issue of implementation size}: those operating in the continuous-wave regime~\cite{Morgenthaler_1958_velocity,Kunz_1980_plane,Fante_1971_transmission,Li_2023_TIR} would imply impractically large device sizes, while those operating in the pulse-wave regime~\cite{Felsen_1970_wave,Mendoncca_2003_temporalsplitter,Biancalana_2007_dynamics,Bacot_2016_timereversal,Akbarzadeh_2018_inverse,Shlivinski_2018_BF,Deck_2018_wave,Pena_2020_tempcoating,Silbiger_2023_filter} would either require complex synchronization (see Sec.~\ref{app:size_problem} in~\cite{supp_mat_pdf}) or preclude extension to the continuous-wave regime.

We provide here a solution to the aforementioned issue of implementation size in aperiodic space-time systems. This solution is based on a transposition of the conventional Fresnel prism concept from two-dimensional space to space-time. The corresponding prism is obtained by periodically re-injecting either a copy of the target modulation or a judiciously adapted version of that modulation into the input of the structure. As the conventional Fresnel prism, the resulting space-time prism can provide drastic device size -- and hence also weight and cost -- reduction.

\section{Space-Time Fresnel Prism Concept}
\begin{figure}[ht!]
    \centering
    \includegraphics[width=8.6cm]{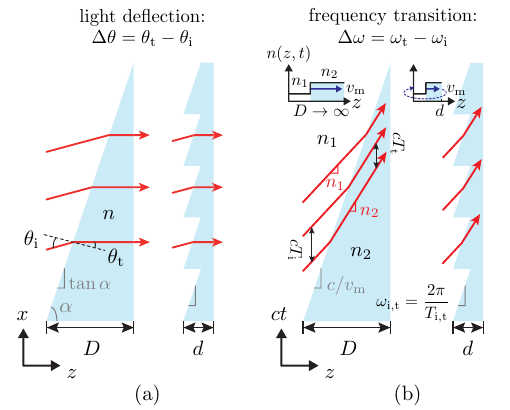}
    \caption{\label{fig:ss_st} Illustration of (a)~conventional (space-space) and (b)~dynamic (space-time) optical prisms (left panels) and their Fresnel counterparts (right panels) with relevant optical structures and light rays. Reflection is assumed to be negligible in both cases.}
\end{figure}

The concept of the proposed Fresnel prism is depicted in Fig.~\ref{fig:ss_st}, with the conventional and space-time prisms being represented in Figs.~\ref{fig:ss_st}(a) and~(b), respectively. Figure~\ref{fig:ss_st}(a) shows the principle of the conventional Fresnel prism~\cite{Saleh_2019_fundamentals,Golub_2006_fresnel}, whereby a bulky prism of refractive index $n$ (left panel) is cut off into a sequence of similar low-profile sub-prisms (right panel) that provides the same light deflection as the original prism with a much smaller form factor ($d\ll D$). 

The space-time Fresnel prism is fundamentally different in nature and serves a totally different purpose while applying the same size reduction principle. The related concept and principle are described in Fig.~\ref{fig:ss_st}(b). The target space-time prism structure consists of a medium with refractive index $n_2$ that is dynamically transformed into a medium with a smaller refractive index $n_1$ by a moving step modulation (left panel with inset). The purpose of that device is to \emph{anharmonically and directionally alter the frequency} of light~\cite{Landauer_1963_velocity,Tsai_1967_wave}, which may have applications as diverse as versatile generation and mixing, magnetless nonreciprocity and absorption, bound-breaking matching and filtering, and classical or quantum processing~\cite{Caloz_2019_spacetime2}. Compared to the conventional prism, the space-time prism replaces the spatial transverse dimension ($x$) [Fig.~\ref{fig:ss_st}(a)] by the time ($ct$) dimension, while maintaining the longitudinal dimension ($z$). It can then be reduced to a Fresnel structure similarly to the conventional prism, as illustrated in the figure (right panel with inset), with the same benefit of drastic device size reduction ($d\ll D$), while providing the same frequency transition operation as the original space-time prism. We restrict here our attention to the case of a single space-time interface, but the proposed concept straightforwardly extends to the cases of a slab or a gradient modulation. 

\section{Two Types of Space-Time \\  Fresnel Prism}
\begin{figure}[ht!]
    \centering
    \includegraphics[width=8.6cm]{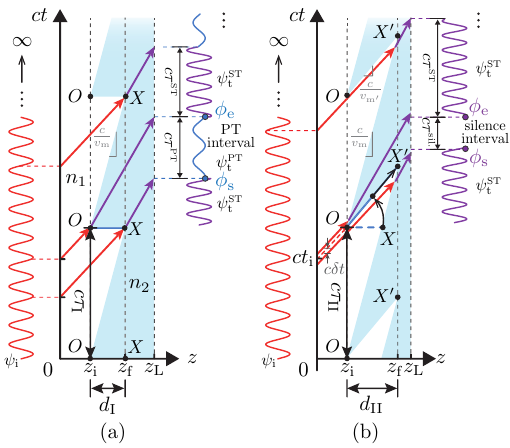}
    \caption{\label{fig:Two_Types} Operation principle of (a)~the Space-Time (ST) Fresnel Prism~I, with Pure-Time (PT) sections [Fig.~\ref{fig:ss_st}(b)], and (b)~the space-time Fresnel Prism~II, with interluminal sections for greater conversion efficiency. The sinusoidal waveforms and related arrows represent the continuous-wave regime; the pulse (possibly multiple-pulse) regime would correspond to restricted sections of these waveforms.}
\end{figure}

The space-time Fresnel prism in Fig.~\ref{fig:ss_st}(b) offers the benefit of simplicity but, as we shall shortly see, it also has some drawbacks, which can be partly mitigated. We shall therefore consider two versions of the space-time Fresnel prism, whose operation principles are depicted in Fig.~\ref{fig:Two_Types} (see Sec.~\ref{app:comparision} in~\cite{supp_mat_pdf}) and which will be detailed later: the initial prism in Fig.~\ref{fig:ss_st}(b), Prism~I, shown in Fig.~\ref{fig:Two_Types}(a), and an improved version of it, Prism~II, shown in Fig.~\ref{fig:Two_Types}(b). 

Prism~I [Fig.~\ref{fig:Two_Types}(a)] involves two types of scattering modulation interfaces: the desired space-time interfaces, which correspond to the oblique sections of the structure and perform the desired space-time frequency conversion, \mbox{$\omega_{\mathrm{t}}^{\mathrm{ST}}=\omega_{\mathrm{i}}(1-\beta n_1)/(1-\beta n_2)$}~\footnote{We may switch from $\omega_{\mathrm{t}}^{\mathrm{ST}}>\omega_{\mathrm{i}}$ (blue shift) to $\omega_{\mathrm{t}}^{\mathrm{ST}}<\omega_{\mathrm{i}}$ (red shift) by making the ratio $n_2/n_1$ larger or smaller than~1, or by making the parameter $\beta$ greater (co-directional modulation) or smaller (contra-directional modulation) than zero. However, changing the ratio $n_2/n_1$ is preferable to changing the velocity regime from $\beta>0$ to $\beta<0$ because $\beta>0$ implies a smaller reflection coefficient~\cite{Caloz_2019_spacetime2}.}, where $\beta=v_{\mathrm{m}}/c$ is the normalized modulation velocity~\cite{Caloz_2019_spacetime2}, over the time intervals $\tau^\mathrm{ST}$, and spurious pure-time interfaces, which correspond to the horizontal sections and produce the undesired pure-time transitions $\omega_{\mathrm{t}}^{\mathrm{PT}}=\omega_{\mathrm{i}}n_1/n_2$~\cite{Caloz_2019_spacetime2} over the intervals $\tau^\mathrm{PT}$ in the output waveform, as shown at the right of the figure. Given its simplicity, this prism might be the solution of choice in the pulse-wave regime, assuming synchronization of the incident pulses with the space-time sections, but the pure-time intervals represent a serious drawback in the continuous-wave regime, with conversion efficiency $\eta_\mathrm{I}=\tau^\mathrm{ST}/(\tau^\mathrm{PT}+\tau^\mathrm{ST})=1-n_2\beta$ (see Sec.~\ref{app:efficiency} in~\cite{supp_mat_pdf}) decreasing to zero in the subluminal limit $v_{\mathrm{m}}\rightarrow c/n_2$.

Prism~II [Fig.~\ref{fig:Two_Types}(b)] is obtained by rotating and translating the vertex points $X$ of Prism~I [Fig.~\ref{fig:Two_Types}(a)] from the vertex points $O$ into the new vertex points $X'$ in the space-time diagram. The rotation, with the angle set between the angles of the incident wave and transmitted wave trajectories, eliminates the pure-time interface sections, while the translation extends the new (interluminal, $c/n_2<v_{\mathrm{m}'}<c/n_1$) interface sections to the modulation final position. As the result of this operation, the PT intervals have been replaced by silence intervals, of duration $\tau^\mathrm{silence}$, as illustrated at the right of the figure. This may a priori suggest that Prism~II offers no benefit over Prism~I, since it also implies an undesired discontinuity in the output waveform. However, its conversion efficiency, \mbox{$\eta_\mathrm{II}=\tau^\mathrm{ST}/(\tau^\mathrm{silence}+\tau^\mathrm{ST})=(1-n_2\beta)/(1-n_1\beta)$} (see Sec.~\ref{app:efficiency} in~\cite{supp_mat_pdf}) is greater, with an enhancement of $\eta_\mathrm{II}/\eta_\mathrm{I}=1/(1-n_1\beta)\overset{v_\mathrm{m}\rightarrow c/n_2}{=}1/(1-n_1/n_2)$ that becomes important for small index contrast.

\section{Spurious Effects Comparison}
At this point, it may be interesting to note the following further differences and similarities between the proposed space-time Fresnel prisms (Fig.~\ref{fig:Two_Types}) and the conventional double-space prism [right panel of Fig.~\ref{fig:ss_st}(a)]. In the latter case, spurious diffraction occurs at the vertices of the structure, whereas no such diffraction occurs in the former case since the related ``rays'', corresponding to space-time trajectories, have fixed incidence and scattering directions [with slopes $n_1$ and $n_2$, see Fig.~\ref{fig:ss_st}(b)], as opposed to displaying a complex angular spectrum spread (see Sec.~\ref{app:spurious} in~\cite{supp_mat_pdf}). On the other hand, the horizontal sections of the conventional prism and Prism~I, although being of different natures, both induce spurious scattering, with different deflection direction in the former case and with the aforementioned pure-time frequency conversion in the latter case (see Sec.~\ref{app:spurious} in~\cite{supp_mat_pdf}); the situation of Prism~II is essentially similar to that of Prism~I insofar as spurious scattering and consequent discontinuity are concerned, except for the replacement of the horizontal pure-time sections by the interluminal space-time sections.

\section{Phase Continuity Design}
As seen above, both Prism~I and Prism~II suffer from spurious output waveform discontinuity, although that issue may be largely mitigated in the latter case. Can one devise a scheme that would completely eliminate the output waveform  -- PT [Fig.~\ref{fig:Two_Types}(a)] or silence [Fig.~\ref{fig:Two_Types}(b)] interval -- discontinuity for applications requiring perfect continuity? We shall try next to address that question. 

Achieving such output waveform continuity would require both ensuring continuity of the phases of the waveform at the starting and ending times of the spurious intervals, i.e., $\phi_{\mathrm{e}}=\phi_{\mathrm{s}}$ in Fig.~\ref{fig:Two_Types}, and interconnecting the corresponding points to close up the spurious gaps. In the case of Prism~II [Fig.~\ref{fig:Two_Types}(b)], the condition $\phi_{\mathrm{e}}=\phi_{\mathrm{s}}$ is automatically satisfied from the fact that the related trajectories (arrows of fixed phase) originate from the same (unique) trajectories impinging on the $O$ vertices before splitting at the input of the device. The re-injection period, $\tau_{\mathrm{II}}$ is related to the device size $d_\mathrm{II}$ as (see Sec.~\ref{app:FP} in~\cite{supp_mat_pdf})
\begin{equation}\label{eq:MFP_dimensions}
    \tau_\mathrm{II}=d_\mathrm{II}(1/v_{\mathrm{m}}-1/v_{\mathrm{m}'}). 
\end{equation}
In contrast, Prism~I [Fig.~\ref{fig:Two_Types}(a)] generally involves $\phi_{\mathrm{e}}\neq\phi_{\mathrm{s}}$, but proper phase continuity, $\phi_{\mathrm{e}}=\phi_{\mathrm{s}}+2\pi q$ ($q\in\mathbb{N}$), can still be obtained under the prism size or temporal period constraint (see Sec.~\ref{app:FP} in~\cite{supp_mat_pdf})
    \begin{equation}\label{eq:FP_dimensions}
        d_\mathrm{I}=q\frac{\lambda_0}{n_1}
        \quad\text{or}\quad
        \tau_\mathrm{I}=\frac{d_\mathrm{I}}{v_{\mathrm{m}}}
        =q\frac{cT_0}{n_1v_{\mathrm{m}}},
    \end{equation}  
where $\lambda_0=2\pi c/\omega_{\mathrm{i}}$ and $T_0=2\pi/\omega_{\mathrm{i}}$ are the free-space incident wavelength and period, respectively, with $\omega_{\mathrm{i}}$ being the incident wave frequency.

\section{Interconnection Schemes}
Now that the phase continuity problem has been resolved, we may consider the interconnection of the desired space-time intervals. Because the related design for Prism~I suffers from the disadvantage of dependency on the incident frequency, according to Eq.~\eqref{eq:FP_dimensions} [$d_\mathrm{I}=2\pi{}cq/(n_1\omega_\mathrm{i})$] and because it is effectively the particular case of Prism~II in the limit $v_\mathrm{m}'\rightarrow\infty$ (Fig.~\ref{fig:Two_Types}), we shall restrict here our attention to the interconnection for Prism~II and defer that for Prism~I to Sec.~\ref{app:delay_FP} in~\cite{supp_mat_pdf}.

\begin{figure}[ht!]
    \centering
    \includegraphics[width=8.6cm]{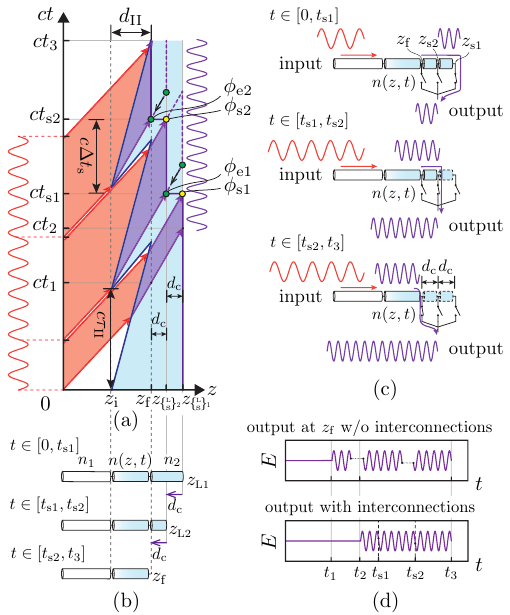}
    \caption{\label{fig:Delay_MFP}Interconnection schemes for suppressing the spurious silence intervals in Prism~II [Fig.~\ref{fig:Two_Types}(b)] with $N$ sub-prisms (here $N=3$). (a)~Space-time diagram required for describing the two schemes, with incidence, space-time and silence zones. (b)~Moving output load scheme. (c)~Switched delay line scheme, involving $N-1$ (here $N-1=2$) transmission line sections between $N$ switches (here $N=3$). (d)~Output waveforms without and with interconnections.}
\end{figure}

We propose here two interconnection schemes, which are depicted in Fig.~\ref{fig:Delay_MFP}. Instead of suppressing the silence intervals $\tau^{\mathrm{silence}}$ in Fig.~\ref{fig:Two_Types}(b), which would be physically impossible, we space-time back-translate, along the corresponding transmitted trajectories, the $\phi_{\mathrm{e}}$-related outputs to the same times as the $\phi_{\mathrm{s}}$-related points, as shown in Fig.~\ref{fig:Delay_MFP}(a). This operation naturally requires some device size extension to the right compared to the initial structure in Fig.~\ref{fig:Two_Types}(b), but the resulting length remains substantially shorter than that of the target prism for a finite number of wave cycles (see Sec.~\ref{app:delay_ratio} in~\cite{supp_mat_pdf}). Conceptually, this scheme could be implemented by mechanically moving an output load backward in space by a distance $d_\mathrm{c}$ at each starting phase time ($t_{\mathrm{s}m}$), as depicted in Fig.~\ref{fig:Delay_MFP}(b). However, this approach would require impractically fast moving parts in a real-world scenario. In contrast, the scheme presented in Fig.~\ref{fig:Delay_MFP}(c), using a sequence of switched delay lines with a common output load, is perfectly realistic and accomplishes exactly the same operation. Figure~\ref{fig:Delay_MFP}(d) plots the output waveforms without and with the two (identical) interconnection schemes, illustrating how that the former continuous-phase sections are interconnected to produce exactly the desired continuous output waveform.

The design of a switched-delay-line Prism~II composed of $N$ sub-prisms [Figs.~\ref{fig:Delay_MFP}(a) and~(c)] involves $N-1$ transmission line sections separated by $N$ switching points, with earlier switching times corresponding to the farther switching positions. In designing the system, the switching points must be placed at the positions (see Sec.~\ref{app:delay_FP} in~\cite{supp_mat_pdf})
\begin{subequations} \label{eq:delay_distance}
    \begin{equation}\label{eq:delay_zlm}
        z_{\mathrm{s}m}=z_\mathrm{f}+(N-m) d_{\mathrm{c}},\quad m=1,2,3,...,N,
    \end{equation}
where $z_{\mathrm{f}}$ is the final position of the modulation and
    \begin{equation}\label{eq:delay_dc}
        d_{\mathrm{c}}=\frac{(c/n_1-c/n_2)(v_{\mathrm{m} '}-v_{\mathrm{m}})}{v_{\mathrm{m}'}(c/n_1-v_{\mathrm{m}})}d_{\mathrm{II}},
    \end{equation}  
\end{subequations} 
while the switching times must be set to the timings
\begin{subequations}\label{eq:delay_tsm}
    \begin{equation}
        t_{\mathrm{s}m}=t_{\mathrm{s}1}+(m-1) \Delta t_{\mathrm{s}},\quad m=1,2,3,...,N,
    \end{equation}
where
    \begin{equation}
        t_{\mathrm{s}1}=\tau_{\mathrm{II}}+\frac{d_{\mathrm{II}}+(N-2)d_{\mathrm{c}} }{c/n_2}
    \end{equation}
is the switching time of the first switch and
    \begin{equation} 
         \Delta t_{\mathrm{s}}=\frac{n_2(v_{\mathrm{m}'}-v_{\mathrm{m}})(c/n_2-v_{\mathrm{m}})}{n_1v_{\mathrm{m}}v_{\mathrm{m}'}(c/n_1-v_{\mathrm{m}})}d_{\mathrm{II}}
    \end{equation}
is the (uniform) switching interval between the switching times.
 \end{subequations}
Note that none of Eqs.~\eqref{eq:MFP_dimensions} and~\eqref{eq:delay_distance}-\eqref{eq:delay_tsm} involves the parameter $\omega_\mathrm{i}$, which reveals that switched-delay-line Prism~II design offers the benefit, compared to its Prism~I counterpart [Eq.~\eqref{eq:FP_dimensions}], of being independent from the incident frequency and hence provides unlimited operational bandwidth within a reasonably weakly dispersive region of modulation media~\footnote{In case of need, one may resort to various dispersion engineering or compensation techniques~\cite{Caloz_PIEEE_10_2011,Fan_2020_constructing} to enhance the bandwidth of the prisms.}.

\section{Simulation Results and Analysis}
\begin{figure}[ht!]
    \centering
    \includegraphics[width=8.6cm]{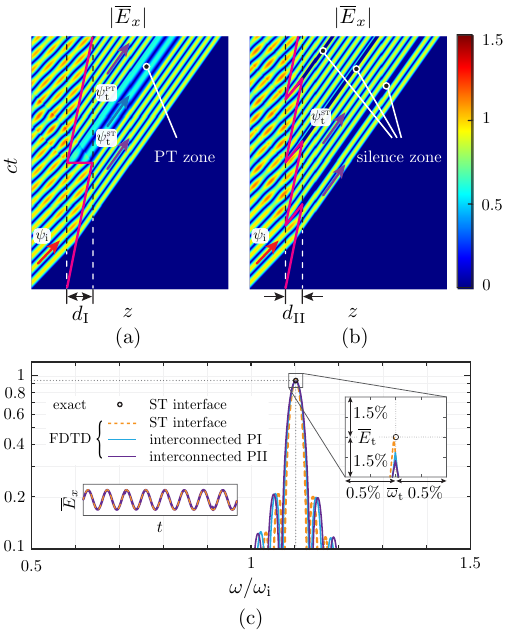}
    \caption{\label{fig:FDTD} Full-wave simulated (FDTD) normalized electric field ($|\overline{E}_{x}|=|E_{x}/E_0|$) for (a)~Prism~I [PI, Fig.~\ref{fig:Two_Types}(a)] and (b)~Prism~II [PII, Fig.~\ref{fig:Two_Types}(b)] with the parameters $\epsilon_1=1.1$, \mbox{$\epsilon_2=2$}, $\mu=1$, $v_{\mathrm{m}}=0.2c$ and $v_{\mathrm{m}'}=0.8 c$, and (c)~corresponding output waveform Fourier transforms, for the cases without and with interconnections (Fig.~\ref{fig:Delay_MFP} for Prism~II and Sec.~\ref{app:delay_FP} in~\cite{supp_mat_pdf} for Prism~I), and including comparison to the target space-time interface [left panel of Fig.~\ref{fig:ss_st}(b)]. The left bottom inset compares the time-domain output waveforms for interconnected Prism~II and the target interface.}
\end{figure}

Figure~\ref{fig:FDTD} provides an illustration and validation of the proposed space-time Fresnel prisms, based on full-wave Finite-Difference Time-Domain (FDTD) simulations~\cite{Taflove_2005_FDTD} utilizing the generalized Yee cell scheme recently reported in~\cite{Deck_2023_yeecell,Bahrami_2023_FDTD} (see Sec.~\ref{app:fdtd} in~\cite{supp_mat_pdf} and animations in~\cite{supp_mat_animation}). We selected here as the excitation a traveling-wave harmonic plane wave with electric field $E_{\mathrm{i}}=E_0\exp[-i(\omega_{\mathrm{i}}t-k_1z)]$ that is injected into the structure at the space-time point $(z,ct)=(0,0)$~\footnote{Although we selected a harmonic plane wave excitation, the forthcoming discussion straightforwardly extends to (modulated or unmodulated) pulse excitations of different shapes.}. Figures~\ref{fig:FDTD}(a) and (b) plot the space-time evolution of scattering for Prism~I and Prism~II, respectively, in the form of space-time diagrams that were obtained by stacking in time successive spatial waveforms obtained by the FDTD simulation. These results qualitatively demonstrate the operation of the prism, with expected (here blue-shifted) frequency transitions, spurious pure-time (here red-shifted frequency) intervals for Prism~I [Fig.~\ref{fig:Two_Types}(a)] and silence intervals [Fig.~\ref{fig:Two_Types}(b)] for Prism~II. The slight standing-wave pattern observed in the first medium is due to small reflection near the space-time regions ($\overline{E}_{\mathrm{r}}=(n_1-n_2)(1-\beta n_1)/[(n_1+n_2)(1+\beta n_1)]$~\cite{Caloz_2019_spacetime2}, with here $|\overline{E}_{\mathrm{r}}|=0.097$~\footnote{The reflection (although already minimal) might be completely suppressed via either impedance matching (\mbox{$\eta_1=\eta_2$}) or oblique injection at the Brewster angle~\cite{Kunz_1980_plane}}.). 

Figure~\ref{fig:FDTD}(c) completes the qualitative results of Figs.~\ref{fig:FDTD}(a) and (b) by providing quantitative validation in terms of Fourier-transformed field distributions. The results for the interconnected designs of both Prisms~I and~II show excellent agreement with theory ($\overline{\omega}_{\mathrm{t}}=\omega_{\mathrm{t}}^{\mathrm{ST}}/\omega_{\mathrm{i}}=(1-\beta n_1)/(1-\beta n_2)$ and $\overline{E}_{\mathrm{t}}=2 n_1(1-\beta n_1)/[(n_1+n_2)(1-\beta n_2)]$~\cite{Caloz_2019_spacetime2}, with here $\overline{\omega}_{\mathrm{t}}=1.102$ and $\overline{E}_{\mathrm{t}}=0.938$) with frequency and magnitude errors of less than $0.5\%$ and $1.5\%$, respectively. This minor discrepancy is due to a combination of the following unaccounted-for effects: i)~local multiple scattering in vertex regions (see Sec.~\ref{app:spurious} in~\cite{supp_mat_pdf}), ii)~finite duration of the testing continuous wave and resulting windowed fast-Fourier transform, and iii)~FDTD discretization. 

\begin{figure}[ht!]
    \centering
    \includegraphics[width=8.6cm]{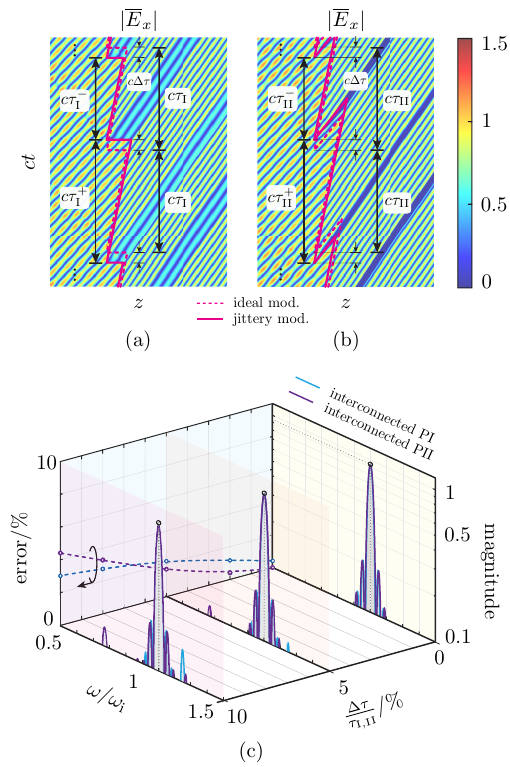}
    \caption{\label{fig:Jitter} Full-wave simulated (FDTD) normalized electric field ($|\overline{E}_{x}|=|E_{x}/E_0|$) illustrating the effect of modulation jitter for (a)~Prism~I and (b)~Prism~II, with the same parameters as Fig.~\ref{fig:FDTD}, for a modulation jitter of $\Delta\tau/\tau_{\mathrm{I,II}}=10\%$, and (c)~corresponding output (after interconnection) spectra and jitter field amplitude error ($|\overline{E}_{\mathrm{t}}^{\mathrm{jittery}}-\overline{E}_{\mathrm{t}}^{\mathrm{exact}}|/\overline{E}_{\mathrm{t}}^{\mathrm{exact}}$, exact: target interface) for modulation jitters of $\Delta\tau/\tau_{\mathrm{I,II}}=0\%$, $5\%$ and $10\%$.}
\end{figure}

The simulation results in Fig.~\ref{fig:FDTD} showcase the capability of synchronized Fresnel prisms to emulate a moving target interface. However, factors such as noise, temperature fluctuations and interference with other devices, may introduce jitter errors~\cite{Li_2007_jitter} in a real-life implementation. Figure~\ref{fig:Jitter} presents simulation results that assess the effect of such jittering in the two proposed prisms, with Figs.~\ref{fig:Jitter}(a) and (b) showing the jittery waveforms of Prism~I and Prism~II, respectively, and Fig.~\ref{fig:Jitter}(c) showing related output spectra and errors. It considers the worst-case scenario of systematic modulation jitter mismatch (periodic advance-delay sequence) in the modulation period, with jittering of up to $10\%$. The corresponding field amplitude error does not exceed $5\%$, which is negligible in typical applications.

\section{Conclusion and Discussion}
We have presented the concept of space-time Fresnel prisms to address the challenge of size reduction in anharmonic and nonreciprocal frequency conversion provided by aperiodic-modulation space-time systems. Such structures may be implemented in various technologies, including switched transmission lines~\cite{Ran_2015_realizingDoppler,Bahrami_2021_Pulse,Moussa_2023_observation}, nonlinear anisotropic perturbations~\cite{Philbin_2008_fiber,Sloan_2022_twophoton}, semiconductor free-carrier injection~\cite{Gaafar_2019_front}, and various other acoustic, electronic, mechanical, plasmonic and optical modulations~\cite{Caloz_GSTEMs}, with the usual single modulation interface being replaced by two independent-velocity and
\mbox{time-periodic} modulation interfaces. The proposed Fresnel prism structures, whose uniform (flat) space-time interfaces provide uniform space-time frequency transitions, may be readily extended to Fresnel lens structures, with accelerated (curved) space-time interfaces~\cite{Bahrami_2023_FDTD} providing unprecedented type of space-time chirping functionalities.

\begin{acknowledgements}
    Z. Li acknowledges financial support from the China Scholarship Council (CSC, No. 202106280230).
\end{acknowledgements}  

\bibliography{Reference}

\begin{thebibliography}{64}%
\makeatletter
\providecommand \@ifxundefined [1]{%
 \@ifx{#1\undefined}
}%
\providecommand \@ifnum [1]{%
 \ifnum #1\expandafter \@firstoftwo
 \else \expandafter \@secondoftwo
 \fi
}%
\providecommand \@ifx [1]{%
 \ifx #1\expandafter \@firstoftwo
 \else \expandafter \@secondoftwo
 \fi
}%
\providecommand \natexlab [1]{#1}%
\providecommand \enquote  [1]{``#1''}%
\providecommand \bibnamefont  [1]{#1}%
\providecommand \bibfnamefont [1]{#1}%
\providecommand \citenamefont [1]{#1}%
\providecommand \href@noop [0]{\@secondoftwo}%
\providecommand \href [0]{\begingroup \@sanitize@url \@href}%
\providecommand \@href[1]{\@@startlink{#1}\@@href}%
\providecommand \@@href[1]{\endgroup#1\@@endlink}%
\providecommand \@sanitize@url [0]{\catcode `\\12\catcode `\$12\catcode `\&12\catcode `\#12\catcode `\^12\catcode `\_12\catcode `\%12\relax}%
\providecommand \@@startlink[1]{}%
\providecommand \@@endlink[0]{}%
\providecommand \url  [0]{\begingroup\@sanitize@url \@url }%
\providecommand \@url [1]{\endgroup\@href {#1}{\urlprefix }}%
\providecommand \urlprefix  [0]{URL }%
\providecommand \Eprint [0]{\href }%
\providecommand \doibase [0]{https://doi.org/}%
\providecommand \selectlanguage [0]{\@gobble}%
\providecommand \bibinfo  [0]{\@secondoftwo}%
\providecommand \bibfield  [0]{\@secondoftwo}%
\providecommand \translation [1]{[#1]}%
\providecommand \BibitemOpen [0]{}%
\providecommand \bibitemStop [0]{}%
\providecommand \bibitemNoStop [0]{.\EOS\space}%
\providecommand \EOS [0]{\spacefactor3000\relax}%
\providecommand \BibitemShut  [1]{\csname bibitem#1\endcsname}%
\let\auto@bib@innerbib\@empty
\bibitem [{\citenamefont {Lurie}(2007)}]{Lurie_2007_mathematical}%
  \BibitemOpen
  \bibfield  {author} {\bibinfo {author} {\bibfnamefont {K.~A.}\ \bibnamefont {Lurie}},\ }\href@noop {} {\emph {\bibinfo {title} {An {I}ntroduction to the {M}athematical {T}heory of {D}ynamic {M}aterials}}}\ (\bibinfo  {publisher} {Springer},\ \bibinfo {year} {2007})\BibitemShut {NoStop}%
\bibitem [{\citenamefont {Caloz}\ and\ \citenamefont {Deck-L{\'e}ger}(2019)}]{Caloz_2019_spacetime1}%
  \BibitemOpen
  \bibfield  {author} {\bibinfo {author} {\bibfnamefont {C.}~\bibnamefont {Caloz}}\ and\ \bibinfo {author} {\bibfnamefont {Z.-L.}\ \bibnamefont {Deck-L{\'e}ger}},\ }\bibfield  {title} {\bibinfo {title} {Spacetime metamaterials—{P}art {I}: General concepts},\ }\href {https://doi.org/10.1109/TAP.2019.2944225} {\bibfield  {journal} {\bibinfo  {journal} {IEEE Trans. Antennas Propag.}\ }\textbf {\bibinfo {volume} {68}},\ \bibinfo {pages} {1569} (\bibinfo {year} {2019})}\BibitemShut {NoStop}%
\bibitem [{\citenamefont {{}Caloz}\ and\ \citenamefont {Deck-L{\'e}ger}(2019)}]{Caloz_2019_spacetime2}%
  \BibitemOpen
  \bibfield  {author} {\bibinfo {author} {\bibfnamefont {C.}~\bibnamefont {{}Caloz}}\ and\ \bibinfo {author} {\bibfnamefont {Z.-L.}\ \bibnamefont {Deck-L{\'e}ger}},\ }\bibfield  {title} {\bibinfo {title} {Spacetime metamaterials—{P}art {II}: Theory and applications},\ }\href {https://doi.org/10.1109/TAP.2019.2944216} {\bibfield  {journal} {\bibinfo  {journal} {IEEE Trans. Antennas Propag.}\ }\textbf {\bibinfo {volume} {68}},\ \bibinfo {pages} {1583} (\bibinfo {year} {2019})}\BibitemShut {NoStop}%
\bibitem [{\citenamefont {Engheta}(2021)}]{Engheta_2021_metamaterials}%
  \BibitemOpen
  \bibfield  {author} {\bibinfo {author} {\bibfnamefont {N.}~\bibnamefont {Engheta}},\ }\bibfield  {title} {\bibinfo {title} {Metamaterials with high degrees of freedom: space, time, and more},\ }\href {https://doi.org/10.1515/nanoph-2020-0414} {\bibfield  {journal} {\bibinfo  {journal} {Nanophotonics}\ }\textbf {\bibinfo {volume} {10}},\ \bibinfo {pages} {639} (\bibinfo {year} {2021})}\BibitemShut {NoStop}%
\bibitem [{\citenamefont {Caloz}\ \emph {et~al.}(2022)\citenamefont {Caloz}, \citenamefont {Deck-L{\'e}ger}, \citenamefont {Bahrami}, \citenamefont {Vicente},\ and\ \citenamefont {Li}}]{Caloz_GSTEMs}%
  \BibitemOpen
  \bibfield  {author} {\bibinfo {author} {\bibfnamefont {C.}~\bibnamefont {Caloz}}, \bibinfo {author} {\bibfnamefont {Z.-L.}\ \bibnamefont {Deck-L{\'e}ger}}, \bibinfo {author} {\bibfnamefont {A.}~\bibnamefont {Bahrami}}, \bibinfo {author} {\bibfnamefont {O.~C.}\ \bibnamefont {Vicente}},\ and\ \bibinfo {author} {\bibfnamefont {Z.}~\bibnamefont {Li}},\ }\bibfield  {title} {\bibinfo {title} {Generalized space-time engineered modulation ({GSTEM}) metamaterials: A global and extended perspective.},\ }\href {https://doi.org/10.1109/MAP.2022.3216773} {\bibfield  {journal} {\bibinfo  {journal} {IEEE Antennas Propag. Mag.}\ ,\ \bibinfo {pages} {2}} (\bibinfo {year} {2022})}\BibitemShut {NoStop}%
\bibitem [{\citenamefont {Engheta}(2023)}]{Engheta_2023_fourdimensional}%
  \BibitemOpen
  \bibfield  {author} {\bibinfo {author} {\bibfnamefont {N.}~\bibnamefont {Engheta}},\ }\bibfield  {title} {\bibinfo {title} {Four-dimensional optics using time-varying metamaterials},\ }\href {https://doi.org/10.1126/science.adf1094} {\bibfield  {journal} {\bibinfo  {journal} {Science}\ }\textbf {\bibinfo {volume} {379}},\ \bibinfo {pages} {1190} (\bibinfo {year} {2023})}\BibitemShut {NoStop}%
\bibitem [{\citenamefont {Yin}\ \emph {et~al.}(2023)\citenamefont {Yin}, \citenamefont {Galiffi}, \citenamefont {Xu},\ and\ \citenamefont {Alù}}]{Alu_2023_scattaringtemporal}%
  \BibitemOpen
  \bibfield  {author} {\bibinfo {author} {\bibfnamefont {S.}~\bibnamefont {Yin}}, \bibinfo {author} {\bibfnamefont {E.}~\bibnamefont {Galiffi}}, \bibinfo {author} {\bibfnamefont {G.}~\bibnamefont {Xu}},\ and\ \bibinfo {author} {\bibfnamefont {A.}~\bibnamefont {Alù}},\ }\bibfield  {title} {\bibinfo {title} {Scattering at temporal interfaces: An overview from an antennas and propagation engineering perspective.},\ }\href {https://doi.org/10.1109/MAP.2023.3254486} {\bibfield  {journal} {\bibinfo  {journal} {IEEE Antennas Propag. Mag.}\ ,\ \bibinfo {pages} {2}} (\bibinfo {year} {2023})}\BibitemShut {NoStop}%
\bibitem [{\citenamefont {Yu}\ and\ \citenamefont {Fan}(2009)}]{Yu_2009_opticalisolation}%
  \BibitemOpen
  \bibfield  {author} {\bibinfo {author} {\bibfnamefont {Z.}~\bibnamefont {Yu}}\ and\ \bibinfo {author} {\bibfnamefont {S.}~\bibnamefont {Fan}},\ }\bibfield  {title} {\bibinfo {title} {Complete optical isolation created by indirect interband photonic transitions},\ }\href {https://doi.org/10.1038/nphoton.2008.273} {\bibfield  {journal} {\bibinfo  {journal} {Nat. Photonics}\ }\textbf {\bibinfo {volume} {3}},\ \bibinfo {pages} {91} (\bibinfo {year} {2009})}\BibitemShut {NoStop}%
\bibitem [{\citenamefont {Estep}\ \emph {et~al.}(2014)\citenamefont {Estep}, \citenamefont {Sounas}, \citenamefont {Soric},\ and\ \citenamefont {Alù}}]{Estep_2014_nonreciprocity}%
  \BibitemOpen
  \bibfield  {author} {\bibinfo {author} {\bibfnamefont {N.~A.}\ \bibnamefont {Estep}}, \bibinfo {author} {\bibfnamefont {D.~L.}\ \bibnamefont {Sounas}}, \bibinfo {author} {\bibfnamefont {J.}~\bibnamefont {Soric}},\ and\ \bibinfo {author} {\bibfnamefont {A.}~\bibnamefont {Alù}},\ }\bibfield  {title} {\bibinfo {title} {Magnetic-free non-reciprocity and isolation based on parametrically modulated coupled-resonator loops},\ }\href {https://doi.org/10.1038/nphys3134} {\bibfield  {journal} {\bibinfo  {journal} {Nat. Phys.}\ }\textbf {\bibinfo {volume} {10}},\ \bibinfo {pages} {923} (\bibinfo {year} {2014})}\BibitemShut {NoStop}%
\bibitem [{\citenamefont {Taravati}\ \emph {et~al.}(2017)\citenamefont {Taravati}, \citenamefont {Chamanara},\ and\ \citenamefont {Caloz}}]{Taravati_2017_nonreciprocal}%
  \BibitemOpen
  \bibfield  {author} {\bibinfo {author} {\bibfnamefont {S.}~\bibnamefont {Taravati}}, \bibinfo {author} {\bibfnamefont {N.}~\bibnamefont {Chamanara}},\ and\ \bibinfo {author} {\bibfnamefont {C.}~\bibnamefont {Caloz}},\ }\bibfield  {title} {\bibinfo {title} {Nonreciprocal electromagnetic scattering from a periodically space-time modulated slab and application to a quasisonic isolator},\ }\href {https://doi.org/10.1103/PhysRevB.96.165144} {\bibfield  {journal} {\bibinfo  {journal} {Phys. Rev. B}\ }\textbf {\bibinfo {volume} {96}},\ \bibinfo {pages} {165144} (\bibinfo {year} {2017})}\BibitemShut {NoStop}%
\bibitem [{\citenamefont {Li}\ \emph {et~al.}(2018)\citenamefont {Li}, \citenamefont {Kottos},\ and\ \citenamefont {Shapiro}}]{Li_2018_floquet}%
  \BibitemOpen
  \bibfield  {author} {\bibinfo {author} {\bibfnamefont {H.}~\bibnamefont {Li}}, \bibinfo {author} {\bibfnamefont {T.}~\bibnamefont {Kottos}},\ and\ \bibinfo {author} {\bibfnamefont {B.}~\bibnamefont {Shapiro}},\ }\bibfield  {title} {\bibinfo {title} {Floquet-network theory of nonreciprocal transport},\ }\href {https://doi.org/10.1103/PhysRevApplied.9.044031} {\bibfield  {journal} {\bibinfo  {journal} {Phys. Rev. Appl.}\ }\textbf {\bibinfo {volume} {9}},\ \bibinfo {pages} {044031} (\bibinfo {year} {2018})}\BibitemShut {NoStop}%
\bibitem [{\citenamefont {Bacot}\ \emph {et~al.}(2016)\citenamefont {Bacot}, \citenamefont {Labousse}, \citenamefont {Eddi}, \citenamefont {Fink},\ and\ \citenamefont {Fort}}]{Bacot_2016_timereversal}%
  \BibitemOpen
  \bibfield  {author} {\bibinfo {author} {\bibfnamefont {V.}~\bibnamefont {Bacot}}, \bibinfo {author} {\bibfnamefont {M.}~\bibnamefont {Labousse}}, \bibinfo {author} {\bibfnamefont {A.}~\bibnamefont {Eddi}}, \bibinfo {author} {\bibfnamefont {M.}~\bibnamefont {Fink}},\ and\ \bibinfo {author} {\bibfnamefont {E.}~\bibnamefont {Fort}},\ }\bibfield  {title} {\bibinfo {title} {Time reversal and holography with spacetime transformations},\ }\href {https://doi.org/10.1038/nphys3810} {\bibfield  {journal} {\bibinfo  {journal} {Nat. Phys.}\ }\textbf {\bibinfo {volume} {12}},\ \bibinfo {pages} {972} (\bibinfo {year} {2016})}\BibitemShut {NoStop}%
\bibitem [{\citenamefont {Deck-L{\'e}ger}\ \emph {et~al.}(2018)\citenamefont {Deck-L{\'e}ger}, \citenamefont {Akbarzadeh},\ and\ \citenamefont {Caloz}}]{Deck_2018_wave}%
  \BibitemOpen
  \bibfield  {author} {\bibinfo {author} {\bibfnamefont {Z.-L.}\ \bibnamefont {Deck-L{\'e}ger}}, \bibinfo {author} {\bibfnamefont {A.}~\bibnamefont {Akbarzadeh}},\ and\ \bibinfo {author} {\bibfnamefont {C.}~\bibnamefont {Caloz}},\ }\bibfield  {title} {\bibinfo {title} {Wave deflection and shifted refocusing in a medium modulated by a superluminal rectangular pulse},\ }\href {https://doi.org/10.1103/PhysRevB.97.104305} {\bibfield  {journal} {\bibinfo  {journal} {Phys. Rev. B}\ }\textbf {\bibinfo {volume} {97}},\ \bibinfo {pages} {104305} (\bibinfo {year} {2018})}\BibitemShut {NoStop}%
\bibitem [{\citenamefont {Akbarzadeh}\ \emph {et~al.}(2018)\citenamefont {Akbarzadeh}, \citenamefont {Chamanara},\ and\ \citenamefont {Caloz}}]{Akbarzadeh_2018_inverse}%
  \BibitemOpen
  \bibfield  {author} {\bibinfo {author} {\bibfnamefont {A.}~\bibnamefont {Akbarzadeh}}, \bibinfo {author} {\bibfnamefont {N.}~\bibnamefont {Chamanara}},\ and\ \bibinfo {author} {\bibfnamefont {C.}~\bibnamefont {Caloz}},\ }\bibfield  {title} {\bibinfo {title} {Inverse prism based on temporal discontinuity and spatial dispersion},\ }\href {https://doi.org/10.1364/OL.43.003297} {\bibfield  {journal} {\bibinfo  {journal} {Opt. Lett.}\ }\textbf {\bibinfo {volume} {43}},\ \bibinfo {pages} {3297} (\bibinfo {year} {2018})}\BibitemShut {NoStop}%
\bibitem [{\citenamefont {Pacheco-Pe{\~n}a}\ and\ \citenamefont {Engheta}(2020)}]{Pena_2020_tempcoating}%
  \BibitemOpen
  \bibfield  {author} {\bibinfo {author} {\bibfnamefont {V.}~\bibnamefont {Pacheco-Pe{\~n}a}}\ and\ \bibinfo {author} {\bibfnamefont {N.}~\bibnamefont {Engheta}},\ }\bibfield  {title} {\bibinfo {title} {Antireflection temporal coatings},\ }\href {https://doi.org/10.1364/OPTICA.381175} {\bibfield  {journal} {\bibinfo  {journal} {Optica}\ }\textbf {\bibinfo {volume} {7}},\ \bibinfo {pages} {323} (\bibinfo {year} {2020})}\BibitemShut {NoStop}%
\bibitem [{\citenamefont {Shlivinski}\ and\ \citenamefont {Hadad}(2018)}]{Shlivinski_2018_BF}%
  \BibitemOpen
  \bibfield  {author} {\bibinfo {author} {\bibfnamefont {A.}~\bibnamefont {Shlivinski}}\ and\ \bibinfo {author} {\bibfnamefont {Y.}~\bibnamefont {Hadad}},\ }\bibfield  {title} {\bibinfo {title} {Beyond the {B}ode-{F}ano bound: Wideband impedance matching for short pulses using temporal switching of transmission-line parameters},\ }\href {https://doi.org/10.1103/PhysRevLett.121.204301} {\bibfield  {journal} {\bibinfo  {journal} {Phys. Rev. Lett.}\ }\textbf {\bibinfo {volume} {121}},\ \bibinfo {pages} {204301} (\bibinfo {year} {2018})}\BibitemShut {NoStop}%
\bibitem [{\citenamefont {Li}\ \emph {et~al.}(2019)\citenamefont {Li}, \citenamefont {Mekawy},\ and\ \citenamefont {Al{\`u}}}]{Li_2019_beyondChu}%
  \BibitemOpen
  \bibfield  {author} {\bibinfo {author} {\bibfnamefont {H.}~\bibnamefont {Li}}, \bibinfo {author} {\bibfnamefont {A.}~\bibnamefont {Mekawy}},\ and\ \bibinfo {author} {\bibfnamefont {A.}~\bibnamefont {Al{\`u}}},\ }\bibfield  {title} {\bibinfo {title} {Beyond {C}hu’s limit with {F}loquet impedance matching},\ }\href {https://doi.org/10.1103/PhysRevLett.123.164102} {\bibfield  {journal} {\bibinfo  {journal} {Phys. Rev. Lett.}\ }\textbf {\bibinfo {volume} {123}},\ \bibinfo {pages} {164102} (\bibinfo {year} {2019})}\BibitemShut {NoStop}%
\bibitem [{\citenamefont {Deck-L{\'e}ger}\ \emph {et~al.}(2019)\citenamefont {Deck-L{\'e}ger}, \citenamefont {Chamanara}, \citenamefont {Skorobogatiy}, \citenamefont {Silveirinha},\ and\ \citenamefont {Caloz}}]{Deck_2019_uniform}%
  \BibitemOpen
  \bibfield  {author} {\bibinfo {author} {\bibfnamefont {Z.-L.}\ \bibnamefont {Deck-L{\'e}ger}}, \bibinfo {author} {\bibfnamefont {N.}~\bibnamefont {Chamanara}}, \bibinfo {author} {\bibfnamefont {M.}~\bibnamefont {Skorobogatiy}}, \bibinfo {author} {\bibfnamefont {M.~G.}\ \bibnamefont {Silveirinha}},\ and\ \bibinfo {author} {\bibfnamefont {C.}~\bibnamefont {Caloz}},\ }\bibfield  {title} {\bibinfo {title} {Uniform-velocity spacetime crystals},\ }\href {https://doi.org/10.1117/1.AP.1.5.056002} {\bibfield  {journal} {\bibinfo  {journal} {Adv. Photonics}\ }\textbf {\bibinfo {volume} {1}},\ \bibinfo {pages} {056002} (\bibinfo {year} {2019})}\BibitemShut {NoStop}%
\bibitem [{\citenamefont {Peng}(2022)}]{Peng_2022_topological}%
  \BibitemOpen
  \bibfield  {author} {\bibinfo {author} {\bibfnamefont {Y.}~\bibnamefont {Peng}},\ }\bibfield  {title} {\bibinfo {title} {Topological space-time crystal},\ }\href {https://doi.org/10.1103/PhysRevLett.128.186802} {\bibfield  {journal} {\bibinfo  {journal} {Phys. Rev. Lett.}\ }\textbf {\bibinfo {volume} {128}},\ \bibinfo {pages} {186802} (\bibinfo {year} {2022})}\BibitemShut {NoStop}%
\bibitem [{\citenamefont {Huidobro}\ \emph {et~al.}(2019)\citenamefont {Huidobro}, \citenamefont {Galiffi}, \citenamefont {Guenneau}, \citenamefont {Craster},\ and\ \citenamefont {Pendry}}]{Huidobro_2019_fresnel}%
  \BibitemOpen
  \bibfield  {author} {\bibinfo {author} {\bibfnamefont {P.~A.}\ \bibnamefont {Huidobro}}, \bibinfo {author} {\bibfnamefont {E.}~\bibnamefont {Galiffi}}, \bibinfo {author} {\bibfnamefont {S.}~\bibnamefont {Guenneau}}, \bibinfo {author} {\bibfnamefont {R.~V.}\ \bibnamefont {Craster}},\ and\ \bibinfo {author} {\bibfnamefont {J.~B.}\ \bibnamefont {Pendry}},\ }\bibfield  {title} {\bibinfo {title} {Fresnel drag in space--time-modulated metamaterials},\ }\href {https://doi.org/10.1073/pnas.1915027116} {\bibfield  {journal} {\bibinfo  {journal} {Proc. Natl. Acad. Sci. U.S.A.}\ }\textbf {\bibinfo {volume} {116}},\ \bibinfo {pages} {24943} (\bibinfo {year} {2019})}\BibitemShut {NoStop}%
\bibitem [{\citenamefont {Bahrami}\ \emph {et~al.}(2023{\natexlab{a}})\citenamefont {Bahrami}, \citenamefont {Deck-L{\'e}ger},\ and\ \citenamefont {Caloz}}]{Bahrami_2023_electrodynamics}%
  \BibitemOpen
  \bibfield  {author} {\bibinfo {author} {\bibfnamefont {A.}~\bibnamefont {Bahrami}}, \bibinfo {author} {\bibfnamefont {Z.-L.}\ \bibnamefont {Deck-L{\'e}ger}},\ and\ \bibinfo {author} {\bibfnamefont {C.}~\bibnamefont {Caloz}},\ }\bibfield  {title} {\bibinfo {title} {Electrodynamics of metamaterials formed by accelerated modulation},\ }\href {https://doi.org/10.1103/PhysRevApplied.19.054044} {\bibfield  {journal} {\bibinfo  {journal} {Phys. Rev. Appl.}\ }\textbf {\bibinfo {volume} {19}},\ \bibinfo {pages} {054044} (\bibinfo {year} {2023}{\natexlab{a}})}\BibitemShut {NoStop}%
\bibitem [{\citenamefont {Tien}(1958)}]{Tien_1958_parametric}%
  \BibitemOpen
  \bibfield  {author} {\bibinfo {author} {\bibfnamefont {P.~K.}\ \bibnamefont {Tien}},\ }\bibfield  {title} {\bibinfo {title} {Parametric amplification and frequency mixing in propagating circuits},\ }\href {https://doi.org/10.1063/1.1723440} {\bibfield  {journal} {\bibinfo  {journal} {J. Appl. Phys.}\ }\textbf {\bibinfo {volume} {29}},\ \bibinfo {pages} {1347} (\bibinfo {year} {1958})}\BibitemShut {NoStop}%
\bibitem [{\citenamefont {Cassedy}\ and\ \citenamefont {Oliner}(1963)}]{Cassedy_1963_dispersion1}%
  \BibitemOpen
  \bibfield  {author} {\bibinfo {author} {\bibfnamefont {E.~S.}\ \bibnamefont {Cassedy}}\ and\ \bibinfo {author} {\bibfnamefont {A.~A.}\ \bibnamefont {Oliner}},\ }\bibfield  {title} {\bibinfo {title} {Dispersion relations in time-space periodic media: Part {I}—stable interactions},\ }\href {https://doi.org/10.1109/PROC.1963.2566} {\bibfield  {journal} {\bibinfo  {journal} {Proc. IEEE}\ }\textbf {\bibinfo {volume} {51}},\ \bibinfo {pages} {1342} (\bibinfo {year} {1963})}\BibitemShut {NoStop}%
\bibitem [{\citenamefont {Cassedy}(1967)}]{Cassedy_1967_dispersion2}%
  \BibitemOpen
  \bibfield  {author} {\bibinfo {author} {\bibfnamefont {E.~S.}\ \bibnamefont {Cassedy}},\ }\bibfield  {title} {\bibinfo {title} {Dispersion relations in time-space periodic media: Part {II}—unstable interactions},\ }\href {https://doi.org/10.1109/PROC.1967.5775} {\bibfield  {journal} {\bibinfo  {journal} {Proc. IEEE}\ }\textbf {\bibinfo {volume} {55}},\ \bibinfo {pages} {1154} (\bibinfo {year} {1967})}\BibitemShut {NoStop}%
\bibitem [{\citenamefont {Reed}\ \emph {et~al.}(2003)\citenamefont {Reed}, \citenamefont {Solja\ifmmode \check{c}\else \v{c}\fi{}i\ifmmode~\acute{c}\else \'{c}\fi{}},\ and\ \citenamefont {Joannopoulos}}]{Reed_2003_color}%
  \BibitemOpen
  \bibfield  {author} {\bibinfo {author} {\bibfnamefont {E.~J.}\ \bibnamefont {Reed}}, \bibinfo {author} {\bibfnamefont {M.}~\bibnamefont {Solja\ifmmode \check{c}\else \v{c}\fi{}i\ifmmode~\acute{c}\else \'{c}\fi{}}},\ and\ \bibinfo {author} {\bibfnamefont {J.~D.}\ \bibnamefont {Joannopoulos}},\ }\bibfield  {title} {\bibinfo {title} {Color of shock waves in photonic crystals},\ }\href {https://doi.org/10.1103/PhysRevLett.90.203904} {\bibfield  {journal} {\bibinfo  {journal} {Phys. Rev. Lett.}\ }\textbf {\bibinfo {volume} {90}},\ \bibinfo {pages} {203904} (\bibinfo {year} {2003})}\BibitemShut {NoStop}%
\bibitem [{\citenamefont {Zurita-S\'anchez}\ \emph {et~al.}(2009)\citenamefont {Zurita-S\'anchez}, \citenamefont {Halevi},\ and\ \citenamefont {Cervantes-Gonz\'alez}}]{Sanchez_2009_tempslab}%
  \BibitemOpen
  \bibfield  {author} {\bibinfo {author} {\bibfnamefont {J.~R.}\ \bibnamefont {Zurita-S\'anchez}}, \bibinfo {author} {\bibfnamefont {P.}~\bibnamefont {Halevi}},\ and\ \bibinfo {author} {\bibfnamefont {J.~C.}\ \bibnamefont {Cervantes-Gonz\'alez}},\ }\bibfield  {title} {\bibinfo {title} {Reflection and transmission of a wave incident on a slab with a time-periodic dielectric function $\epsilon(t)$},\ }\href {https://doi.org/10.1103/PhysRevA.79.053821} {\bibfield  {journal} {\bibinfo  {journal} {Phys. Rev. A}\ }\textbf {\bibinfo {volume} {79}},\ \bibinfo {pages} {053821} (\bibinfo {year} {2009})}\BibitemShut {NoStop}%
\bibitem [{\citenamefont {Mart{\'\i}nez-Romero}\ \emph {et~al.}(2016)\citenamefont {Mart{\'\i}nez-Romero}, \citenamefont {Becerra-Fuentes},\ and\ \citenamefont {Halevi}}]{Martinez_2016_tempcrystal}%
  \BibitemOpen
  \bibfield  {author} {\bibinfo {author} {\bibfnamefont {J.~S.}\ \bibnamefont {Mart{\'\i}nez-Romero}}, \bibinfo {author} {\bibfnamefont {O.~M.}\ \bibnamefont {Becerra-Fuentes}},\ and\ \bibinfo {author} {\bibfnamefont {P.}~\bibnamefont {Halevi}},\ }\bibfield  {title} {\bibinfo {title} {Temporal photonic crystals with modulations of both permittivity and permeability},\ }\href {https://doi.org/10.1103/PhysRevA.93.063813} {\bibfield  {journal} {\bibinfo  {journal} {Phys. Rev. A}\ }\textbf {\bibinfo {volume} {93}},\ \bibinfo {pages} {063813} (\bibinfo {year} {2016})}\BibitemShut {NoStop}%
\bibitem [{\citenamefont {Mirmoosa}\ \emph {et~al.}(2019)\citenamefont {Mirmoosa}, \citenamefont {Ptitcyn}, \citenamefont {Asadchy},\ and\ \citenamefont {Tretyakov}}]{Mirmoosa_2019_timereactive}%
  \BibitemOpen
  \bibfield  {author} {\bibinfo {author} {\bibfnamefont {M.~S.}\ \bibnamefont {Mirmoosa}}, \bibinfo {author} {\bibfnamefont {G.~A.}\ \bibnamefont {Ptitcyn}}, \bibinfo {author} {\bibfnamefont {V.~S.}\ \bibnamefont {Asadchy}},\ and\ \bibinfo {author} {\bibfnamefont {S.~A.}\ \bibnamefont {Tretyakov}},\ }\bibfield  {title} {\bibinfo {title} {Time-varying reactive elements for extreme accumulation of electromagnetic energy},\ }\href {https://doi.org/10.1103/PhysRevApplied.11.014024} {\bibfield  {journal} {\bibinfo  {journal} {Phys. Rev. Appl.}\ }\textbf {\bibinfo {volume} {11}},\ \bibinfo {pages} {014024} (\bibinfo {year} {2019})}\BibitemShut {NoStop}%
\bibitem [{\citenamefont {Galiffi}\ \emph {et~al.}(2019)\citenamefont {Galiffi}, \citenamefont {Huidobro},\ and\ \citenamefont {Pendry}}]{Galiffi_2019_broadband}%
  \BibitemOpen
  \bibfield  {author} {\bibinfo {author} {\bibfnamefont {E.}~\bibnamefont {Galiffi}}, \bibinfo {author} {\bibfnamefont {P.~A.}\ \bibnamefont {Huidobro}},\ and\ \bibinfo {author} {\bibfnamefont {J.~B.}\ \bibnamefont {Pendry}},\ }\bibfield  {title} {\bibinfo {title} {Broadband nonreciprocal amplification in luminal metamaterials},\ }\href {https://doi.org/10.1103/PhysRevLett.123.206101} {\bibfield  {journal} {\bibinfo  {journal} {Phys. Rev. Lett.}\ }\textbf {\bibinfo {volume} {123}},\ \bibinfo {pages} {206101} (\bibinfo {year} {2019})}\BibitemShut {NoStop}%
\bibitem [{\citenamefont {Pendry}\ \emph {et~al.}(2022)\citenamefont {Pendry}, \citenamefont {Galiffi},\ and\ \citenamefont {Huidobro}}]{Pendry_2022_photon}%
  \BibitemOpen
  \bibfield  {author} {\bibinfo {author} {\bibfnamefont {J.}~\bibnamefont {Pendry}}, \bibinfo {author} {\bibfnamefont {E.}~\bibnamefont {Galiffi}},\ and\ \bibinfo {author} {\bibfnamefont {P.}~\bibnamefont {Huidobro}},\ }\bibfield  {title} {\bibinfo {title} {Photon conservation in trans-luminal metamaterials},\ }\href {https://doi.org/10.1364/OPTICA.462488} {\bibfield  {journal} {\bibinfo  {journal} {Optica}\ }\textbf {\bibinfo {volume} {9}},\ \bibinfo {pages} {724} (\bibinfo {year} {2022})}\BibitemShut {NoStop}%
\bibitem [{\citenamefont {Xu}\ \emph {et~al.}(2022)\citenamefont {Xu}, \citenamefont {Mai},\ and\ \citenamefont {Werner}}]{Xu_2022_tempGTMM}%
  \BibitemOpen
  \bibfield  {author} {\bibinfo {author} {\bibfnamefont {J.}~\bibnamefont {Xu}}, \bibinfo {author} {\bibfnamefont {W.}~\bibnamefont {Mai}},\ and\ \bibinfo {author} {\bibfnamefont {D.~H.}\ \bibnamefont {Werner}},\ }\bibfield  {title} {\bibinfo {title} {Generalized temporal transfer matrix method: a systematic approach to solving electromagnetic wave scattering in temporally stratified structures},\ }\href {https://doi.org/10.1515/nanoph-2021-0715} {\bibfield  {journal} {\bibinfo  {journal} {Nanophotonics}\ }\textbf {\bibinfo {volume} {11}},\ \bibinfo {pages} {1309} (\bibinfo {year} {2022})}\BibitemShut {NoStop}%
\bibitem [{\citenamefont {Morgenthaler}(1958)}]{Morgenthaler_1958_velocity}%
  \BibitemOpen
  \bibfield  {author} {\bibinfo {author} {\bibfnamefont {F.}~\bibnamefont {Morgenthaler}},\ }\bibfield  {title} {\bibinfo {title} {Velocity modulation of electromagnetic waves},\ }\href {https://doi.org/10.1109/TMTT.1958.1124533} {\bibfield  {journal} {\bibinfo  {journal} {IEEE Trans. Microw. Theory Techn.}\ }\textbf {\bibinfo {volume} {6}},\ \bibinfo {pages} {167} (\bibinfo {year} {1958})}\BibitemShut {NoStop}%
\bibitem [{\citenamefont {Felsen}\ and\ \citenamefont {Whitman}(1970)}]{Felsen_1970_wave}%
  \BibitemOpen
  \bibfield  {author} {\bibinfo {author} {\bibfnamefont {L.}~\bibnamefont {Felsen}}\ and\ \bibinfo {author} {\bibfnamefont {G.}~\bibnamefont {Whitman}},\ }\bibfield  {title} {\bibinfo {title} {Wave propagation in time-varying media},\ }\href {https://doi.org/10.1109/TAP.1970.1139657} {\bibfield  {journal} {\bibinfo  {journal} {IEEE Trans. Antennas Propag.}\ }\textbf {\bibinfo {volume} {18}},\ \bibinfo {pages} {242} (\bibinfo {year} {1970})}\BibitemShut {NoStop}%
\bibitem [{\citenamefont {Fante}(1971)}]{Fante_1971_transmission}%
  \BibitemOpen
  \bibfield  {author} {\bibinfo {author} {\bibfnamefont {R.}~\bibnamefont {Fante}},\ }\bibfield  {title} {\bibinfo {title} {Transmission of electromagnetic waves into time-varying media},\ }\href {https://doi.org/10.1109/TAP.1971.1139931} {\bibfield  {journal} {\bibinfo  {journal} {IEEE Trans. Antennas Propag.}\ }\textbf {\bibinfo {volume} {19}},\ \bibinfo {pages} {417} (\bibinfo {year} {1971})}\BibitemShut {NoStop}%
\bibitem [{\citenamefont {Kunz}(1980)}]{Kunz_1980_plane}%
  \BibitemOpen
  \bibfield  {author} {\bibinfo {author} {\bibfnamefont {K.~S.}\ \bibnamefont {Kunz}},\ }\bibfield  {title} {\bibinfo {title} {Plane electromagnetic waves in moving media and reflections from moving interfaces},\ }\href {https://doi.org/10.1063/1.327661} {\bibfield  {journal} {\bibinfo  {journal} {J. Appl. Phys.}\ }\textbf {\bibinfo {volume} {51}},\ \bibinfo {pages} {873} (\bibinfo {year} {1980})}\BibitemShut {NoStop}%
\bibitem [{\citenamefont {Mendon{\c{c}}a}\ \emph {et~al.}(2003)\citenamefont {Mendon{\c{c}}a}, \citenamefont {Martins},\ and\ \citenamefont {Guerreiro}}]{Mendoncca_2003_temporalsplitter}%
  \BibitemOpen
  \bibfield  {author} {\bibinfo {author} {\bibfnamefont {J.~T.}\ \bibnamefont {Mendon{\c{c}}a}}, \bibinfo {author} {\bibfnamefont {A.~M.}\ \bibnamefont {Martins}},\ and\ \bibinfo {author} {\bibfnamefont {A.}~\bibnamefont {Guerreiro}},\ }\bibfield  {title} {\bibinfo {title} {Temporal beam splitter and temporal interference},\ }\href {https://doi.org/10.1103/PhysRevA.68.043801} {\bibfield  {journal} {\bibinfo  {journal} {Phys. Rev. A}\ }\textbf {\bibinfo {volume} {68}},\ \bibinfo {pages} {043801} (\bibinfo {year} {2003})}\BibitemShut {NoStop}%
\bibitem [{\citenamefont {Biancalana}\ \emph {et~al.}(2007)\citenamefont {Biancalana}, \citenamefont {Amann}, \citenamefont {Uskov},\ and\ \citenamefont {O'Reilly}}]{Biancalana_2007_dynamics}%
  \BibitemOpen
  \bibfield  {author} {\bibinfo {author} {\bibfnamefont {F.}~\bibnamefont {Biancalana}}, \bibinfo {author} {\bibfnamefont {A.}~\bibnamefont {Amann}}, \bibinfo {author} {\bibfnamefont {A.~V.}\ \bibnamefont {Uskov}},\ and\ \bibinfo {author} {\bibfnamefont {E.~P.}\ \bibnamefont {O'Reilly}},\ }\bibfield  {title} {\bibinfo {title} {Dynamics of light propagation in spatiotemporal dielectric structures},\ }\href {https://doi.org/10.1103/PhysRevE.75.046607} {\bibfield  {journal} {\bibinfo  {journal} {Phys. Rev. E}\ }\textbf {\bibinfo {volume} {75}},\ \bibinfo {pages} {046607} (\bibinfo {year} {2007})}\BibitemShut {NoStop}%
\bibitem [{\citenamefont {Silbiger}\ and\ \citenamefont {Hadad}(2023)}]{Silbiger_2023_filter}%
  \BibitemOpen
  \bibfield  {author} {\bibinfo {author} {\bibfnamefont {O.}~\bibnamefont {Silbiger}}\ and\ \bibinfo {author} {\bibfnamefont {Y.}~\bibnamefont {Hadad}},\ }\bibfield  {title} {\bibinfo {title} {Optimization-free filter and matched-filter design through spatial and temporal soft switching of the dielectric constant},\ }\href {https://doi.org/10.1103/PhysRevApplied.19.014047} {\bibfield  {journal} {\bibinfo  {journal} {Phys. Rev. Appl.}\ }\textbf {\bibinfo {volume} {19}},\ \bibinfo {pages} {014047} (\bibinfo {year} {2023})}\BibitemShut {NoStop}%
\bibitem [{\citenamefont {Li}\ \emph {et~al.}(2023)\citenamefont {Li}, \citenamefont {Ma}, \citenamefont {Bahrami}, \citenamefont {Deck-L{\'e}ger},\ and\ \citenamefont {Caloz}}]{Li_2023_TIR}%
  \BibitemOpen
  \bibfield  {author} {\bibinfo {author} {\bibfnamefont {Z.}~\bibnamefont {Li}}, \bibinfo {author} {\bibfnamefont {X.}~\bibnamefont {Ma}}, \bibinfo {author} {\bibfnamefont {A.}~\bibnamefont {Bahrami}}, \bibinfo {author} {\bibfnamefont {Z.-L.}\ \bibnamefont {Deck-L{\'e}ger}},\ and\ \bibinfo {author} {\bibfnamefont {C.}~\bibnamefont {Caloz}},\ }\bibfield  {title} {\bibinfo {title} {Generalized total internal reflection at dynamic interfaces},\ }\href {https://doi.org/10.1103/PhysRevB.107.115129} {\bibfield  {journal} {\bibinfo  {journal} {Phys. Rev. B}\ }\textbf {\bibinfo {volume} {107}},\ \bibinfo {pages} {115129} (\bibinfo {year} {2023})}\BibitemShut {NoStop}%
\bibitem [{sup({\natexlab{a}})}]{supp_mat_pdf}%
  \BibitemOpen
  \bibinfo {note} {{S}ee Supplemental Material at [URL will be inserted by publisher] for the derivations, additional figures and extra simulation results}\BibitemShut {NoStop}%
\bibitem [{\citenamefont {Saleh}\ and\ \citenamefont {Teich}(2019)}]{Saleh_2019_fundamentals}%
  \BibitemOpen
  \bibfield  {author} {\bibinfo {author} {\bibfnamefont {B.~E.}\ \bibnamefont {Saleh}}\ and\ \bibinfo {author} {\bibfnamefont {M.~C.}\ \bibnamefont {Teich}},\ }\href@noop {} {\emph {\bibinfo {title} {Fundamentals of {P}hotonics}}}\ (\bibinfo  {publisher} {John Wiley \& {S}ons},\ \bibinfo {year} {2019})\BibitemShut {NoStop}%
\bibitem [{\citenamefont {Golub}(2006)}]{Golub_2006_fresnel}%
  \BibitemOpen
  \bibfield  {author} {\bibinfo {author} {\bibfnamefont {I.}~\bibnamefont {Golub}},\ }\bibfield  {title} {\bibinfo {title} {Fresnel axicon},\ }\href {https://doi.org/10.1364/OL.31.001890} {\bibfield  {journal} {\bibinfo  {journal} {Opt. Lett.}\ }\textbf {\bibinfo {volume} {31}},\ \bibinfo {pages} {1890} (\bibinfo {year} {2006})}\BibitemShut {NoStop}%
\bibitem [{\citenamefont {Landauer}(1963)}]{Landauer_1963_velocity}%
  \BibitemOpen
  \bibfield  {author} {\bibinfo {author} {\bibfnamefont {R.}~\bibnamefont {Landauer}},\ }\bibfield  {title} {\bibinfo {title} {Velocity modulation of propagating waves},\ }\href {https://doi.org/10.1063/1.1729708} {\bibfield  {journal} {\bibinfo  {journal} {J. Appl. Phys.}\ }\textbf {\bibinfo {volume} {34}},\ \bibinfo {pages} {1893} (\bibinfo {year} {1963})}\BibitemShut {NoStop}%
\bibitem [{\citenamefont {Tsai}\ and\ \citenamefont {Auld}(1967)}]{Tsai_1967_wave}%
  \BibitemOpen
  \bibfield  {author} {\bibinfo {author} {\bibfnamefont {C.}~\bibnamefont {Tsai}}\ and\ \bibinfo {author} {\bibfnamefont {B.}~\bibnamefont {Auld}},\ }\bibfield  {title} {\bibinfo {title} {Wave interactions with moving boundaries},\ }\href {https://doi.org/10.1063/1.1709838} {\bibfield  {journal} {\bibinfo  {journal} {J. Appl. Phys.}\ }\textbf {\bibinfo {volume} {38}},\ \bibinfo {pages} {2106} (\bibinfo {year} {1967})}\BibitemShut {NoStop}%
\bibitem [{Note1()}]{Note1}%
  \BibitemOpen
  \bibinfo {note} {We may switch from $\omega _{\protect \mathrm {t}}^{\protect \mathrm {ST}}>\omega _{\protect \mathrm {i}}$ (blue shift) to $\omega _{\protect \mathrm {t}}^{\protect \mathrm {ST}}<\omega _{\protect \mathrm {i}}$ (red shift) by making the ratio $n_2/n_1$ larger or smaller than~1, or by making the parameter $\beta $ greater (co-directional modulation) or smaller (contra-directional modulation) than zero. However, changing the ratio $n_2/n_1$ is preferable to changing the velocity regime from $\beta >0$ to $\beta <0$ because $\beta >0$ implies a smaller reflection coefficient~\cite {Caloz_2019_spacetime2}.}\BibitemShut {Stop}%
\bibitem [{Note2()}]{Note2}%
  \BibitemOpen
  \bibinfo {note} {In case of need, one may resort to various dispersion engineering or compensation techniques~\cite {Caloz_PIEEE_10_2011,Fan_2020_constructing} to enhance the bandwidth of the prisms.}\BibitemShut {Stop}%
\bibitem [{\citenamefont {Taflove}\ and\ \citenamefont {Hagness}(2005)}]{Taflove_2005_FDTD}%
  \BibitemOpen
  \bibfield  {author} {\bibinfo {author} {\bibfnamefont {A.}~\bibnamefont {Taflove}}\ and\ \bibinfo {author} {\bibfnamefont {S.}~\bibnamefont {Hagness}},\ }\href@noop {} {\emph {\bibinfo {title} {Computational {E}lectrodynamics: The {F}inite-{D}ifference {T}ime-{D}omain {M}ethod}}},\ Artech House antennas and propagation library\ (\bibinfo  {publisher} {Artech House},\ \bibinfo {year} {2005})\BibitemShut {NoStop}%
\bibitem [{\citenamefont {Deck-L\'{e}ger}\ \emph {et~al.}(2023)\citenamefont {Deck-L\'{e}ger}, \citenamefont {Bahrami}, \citenamefont {Li},\ and\ \citenamefont {Caloz}}]{Deck_2023_yeecell}%
  \BibitemOpen
  \bibfield  {author} {\bibinfo {author} {\bibfnamefont {Z.-L.}\ \bibnamefont {Deck-L\'{e}ger}}, \bibinfo {author} {\bibfnamefont {A.}~\bibnamefont {Bahrami}}, \bibinfo {author} {\bibfnamefont {Z.}~\bibnamefont {Li}},\ and\ \bibinfo {author} {\bibfnamefont {C.}~\bibnamefont {Caloz}},\ }\bibfield  {title} {\bibinfo {title} {Generalized {FDTD} scheme for the simulation of electromagnetic scattering in moving structures},\ }\href {https://doi.org/10.1364/OE.493099} {\bibfield  {journal} {\bibinfo  {journal} {Opt. Express}\ }\textbf {\bibinfo {volume} {31}},\ \bibinfo {pages} {23214} (\bibinfo {year} {2023})}\BibitemShut {NoStop}%
\bibitem [{\citenamefont {Bahrami}\ \emph {et~al.}(2023{\natexlab{b}})\citenamefont {Bahrami}, \citenamefont {Deck-L{\'e}ger}, \citenamefont {Li},\ and\ \citenamefont {Caloz}}]{Bahrami_2023_FDTD}%
  \BibitemOpen
  \bibfield  {author} {\bibinfo {author} {\bibfnamefont {A.}~\bibnamefont {Bahrami}}, \bibinfo {author} {\bibfnamefont {Z.-L.}\ \bibnamefont {Deck-L{\'e}ger}}, \bibinfo {author} {\bibfnamefont {Z.}~\bibnamefont {Li}},\ and\ \bibinfo {author} {\bibfnamefont {C.}~\bibnamefont {Caloz}},\ }\href@noop {} {\bibinfo {title} {Generalized {FDTD} scheme for moving electromagnetic structures with arbitrary space-time configurations}} (\bibinfo {year} {2023}{\natexlab{b}}),\ \Eprint {https://arxiv.org/abs/arXiv:2306.10035} {arXiv:2306.10035} \BibitemShut {NoStop}%
\bibitem [{sup({\natexlab{b}})}]{supp_mat_animation}%
  \BibitemOpen
  \bibinfo {note} {{S}ee Supplemental Material at [URL will be inserted by publisher] for the FDTD animations}\BibitemShut {NoStop}%
\bibitem [{Note3()}]{Note3}%
  \BibitemOpen
  \bibinfo {note} {Although we selected a harmonic plane wave excitation, the forthcoming discussion straightforwardly extends to (modulated or unmodulated) pulse excitations of different shapes.}\BibitemShut {Stop}%
\bibitem [{Note4()}]{Note4}%
  \BibitemOpen
  \bibinfo {note} {The reflection (although already minimal) might be completely suppressed via either impedance matching (\protect \mbox {$\eta _1=\eta _2$}) or oblique injection at the Brewster angle~\cite {Kunz_1980_plane}}\BibitemShut {NoStop}%
\bibitem [{\citenamefont {Li}(2007)}]{Li_2007_jitter}%
  \BibitemOpen
  \bibfield  {author} {\bibinfo {author} {\bibfnamefont {M.~P.}\ \bibnamefont {Li}},\ }\href@noop {} {\emph {\bibinfo {title} {Jitter, {N}oise, and {S}ignal {I}ntegrity at {H}igh-{S}peed}}}\ (\bibinfo  {publisher} {Pearson Education},\ \bibinfo {year} {2007})\BibitemShut {NoStop}%
\bibitem [{\citenamefont {Ran}\ \emph {et~al.}(2015)\citenamefont {Ran}, \citenamefont {Zhang}, \citenamefont {Chen}, \citenamefont {Fang}, \citenamefont {Zhao}, \citenamefont {Sun},\ and\ \citenamefont {Chen}}]{Ran_2015_realizingDoppler}%
  \BibitemOpen
  \bibfield  {author} {\bibinfo {author} {\bibfnamefont {J.}~\bibnamefont {Ran}}, \bibinfo {author} {\bibfnamefont {Y.}~\bibnamefont {Zhang}}, \bibinfo {author} {\bibfnamefont {X.}~\bibnamefont {Chen}}, \bibinfo {author} {\bibfnamefont {K.}~\bibnamefont {Fang}}, \bibinfo {author} {\bibfnamefont {J.}~\bibnamefont {Zhao}}, \bibinfo {author} {\bibfnamefont {Y.}~\bibnamefont {Sun}},\ and\ \bibinfo {author} {\bibfnamefont {H.}~\bibnamefont {Chen}},\ }\bibfield  {title} {\bibinfo {title} {Realizing tunable inverse and normal {D}oppler shifts in reconfigurable {RF} metamaterials},\ }\href {https://doi.org/10.1038/srep11659} {\bibfield  {journal} {\bibinfo  {journal} {Sci. Rep.}\ }\textbf {\bibinfo {volume} {5}},\ \bibinfo {pages} {11659} (\bibinfo {year} {2015})}\BibitemShut {NoStop}%
\bibitem [{\citenamefont {Bahrami}\ and\ \citenamefont {Caloz}(2021)}]{Bahrami_2021_Pulse}%
  \BibitemOpen
  \bibfield  {author} {\bibinfo {author} {\bibfnamefont {A.}~\bibnamefont {Bahrami}}\ and\ \bibinfo {author} {\bibfnamefont {C.}~\bibnamefont {Caloz}},\ }\bibfield  {title} {\bibinfo {title} {Arbitrary pulse shaping using nonuniform spacetime modulation},\ }in\ \href {https://doi.org/10.1109/APS/URSI47566.2021.9704222} {\emph {\bibinfo {booktitle} {2021 IEEE International Symposium on Antennas and Propagation and USNC-URSI Radio Science Meeting (APS/URSI)}}}\ (\bibinfo {year} {2021})\ pp.\ \bibinfo {pages} {409--410}\BibitemShut {NoStop}%
\bibitem [{\citenamefont {Moussa}\ \emph {et~al.}(2023)\citenamefont {Moussa}, \citenamefont {Xu}, \citenamefont {Yin}, \citenamefont {Galiffi}, \citenamefont {Ra’di},\ and\ \citenamefont {Al{\`u}}}]{Moussa_2023_observation}%
  \BibitemOpen
  \bibfield  {author} {\bibinfo {author} {\bibfnamefont {H.}~\bibnamefont {Moussa}}, \bibinfo {author} {\bibfnamefont {G.}~\bibnamefont {Xu}}, \bibinfo {author} {\bibfnamefont {S.}~\bibnamefont {Yin}}, \bibinfo {author} {\bibfnamefont {E.}~\bibnamefont {Galiffi}}, \bibinfo {author} {\bibfnamefont {Y.}~\bibnamefont {Ra’di}},\ and\ \bibinfo {author} {\bibfnamefont {A.}~\bibnamefont {Al{\`u}}},\ }\bibfield  {title} {\bibinfo {title} {Observation of temporal reflection and broadband frequency translation at photonic time interfaces},\ }\href {https://doi.org/10.1038/s41567-023-01975-y} {\bibfield  {journal} {\bibinfo  {journal} {Nat. Phys.}\ ,\ \bibinfo {pages} {1}} (\bibinfo {year} {2023})}\BibitemShut {NoStop}%
\bibitem [{\citenamefont {Philbin}\ \emph {et~al.}(2008)\citenamefont {Philbin}, \citenamefont {Kuklewicz}, \citenamefont {Robertson}, \citenamefont {Hill}, \citenamefont {König},\ and\ \citenamefont {Leonhardt}}]{Philbin_2008_fiber}%
  \BibitemOpen
  \bibfield  {author} {\bibinfo {author} {\bibfnamefont {T.~G.}\ \bibnamefont {Philbin}}, \bibinfo {author} {\bibfnamefont {C.}~\bibnamefont {Kuklewicz}}, \bibinfo {author} {\bibfnamefont {S.}~\bibnamefont {Robertson}}, \bibinfo {author} {\bibfnamefont {S.}~\bibnamefont {Hill}}, \bibinfo {author} {\bibfnamefont {F.}~\bibnamefont {König}},\ and\ \bibinfo {author} {\bibfnamefont {U.}~\bibnamefont {Leonhardt}},\ }\bibfield  {title} {\bibinfo {title} {Fiber-optical analog of the event horizon},\ }\href {https://doi.org/10.1126/science.1153625} {\bibfield  {journal} {\bibinfo  {journal} {Science}\ }\textbf {\bibinfo {volume} {319}},\ \bibinfo {pages} {1367} (\bibinfo {year} {2008})}\BibitemShut {NoStop}%
\bibitem [{\citenamefont {Sloan}\ \emph {et~al.}(2022)\citenamefont {Sloan}, \citenamefont {Rivera}, \citenamefont {Joannopoulos},\ and\ \citenamefont {Solja{\v{c}}i{\'c}}}]{Sloan_2022_twophoton}%
  \BibitemOpen
  \bibfield  {author} {\bibinfo {author} {\bibfnamefont {J.}~\bibnamefont {Sloan}}, \bibinfo {author} {\bibfnamefont {N.}~\bibnamefont {Rivera}}, \bibinfo {author} {\bibfnamefont {J.~D.}\ \bibnamefont {Joannopoulos}},\ and\ \bibinfo {author} {\bibfnamefont {M.}~\bibnamefont {Solja{\v{c}}i{\'c}}},\ }\bibfield  {title} {\bibinfo {title} {Controlling two-photon emission from superluminal and accelerating index perturbations},\ }\href {https://doi.org/10.1038/s41567-021-01428-4} {\bibfield  {journal} {\bibinfo  {journal} {Nat. Phys.}\ }\textbf {\bibinfo {volume} {18}},\ \bibinfo {pages} {67} (\bibinfo {year} {2022})}\BibitemShut {NoStop}%
\bibitem [{\citenamefont {Gaafar}\ \emph {et~al.}(2019)\citenamefont {Gaafar}, \citenamefont {Baba}, \citenamefont {Eich},\ and\ \citenamefont {Petrov}}]{Gaafar_2019_front}%
  \BibitemOpen
  \bibfield  {author} {\bibinfo {author} {\bibfnamefont {M.~A.}\ \bibnamefont {Gaafar}}, \bibinfo {author} {\bibfnamefont {T.}~\bibnamefont {Baba}}, \bibinfo {author} {\bibfnamefont {M.}~\bibnamefont {Eich}},\ and\ \bibinfo {author} {\bibfnamefont {A.~Y.}\ \bibnamefont {Petrov}},\ }\bibfield  {title} {\bibinfo {title} {Front-induced transitions},\ }\href {https://doi.org/10.1038/s41566-019-0511-6} {\bibfield  {journal} {\bibinfo  {journal} {Nat. Photonics}\ }\textbf {\bibinfo {volume} {13}},\ \bibinfo {pages} {737} (\bibinfo {year} {2019})}\BibitemShut {NoStop}%
\bibitem [{\citenamefont {Caloz}(2011)}]{Caloz_PIEEE_10_2011}%
  \BibitemOpen
  \bibfield  {author} {\bibinfo {author} {\bibfnamefont {C.}~\bibnamefont {Caloz}},\ }\bibfield  {title} {\bibinfo {title} {Metamaterial dispersion engineering concepts and applications},\ }\href {https://doi.org/10.1109/JPROC.2011.2114631} {\bibfield  {journal} {\bibinfo  {journal} {Proc. IEEE}\ }\textbf {\bibinfo {volume} {99}},\ \bibinfo {pages} {1711} (\bibinfo {year} {2011})}\BibitemShut {NoStop}%
\bibitem [{\citenamefont {Fan}\ \emph {et~al.}(2020)\citenamefont {Fan}, \citenamefont {Xiong}, \citenamefont {Peng},\ and\ \citenamefont {Wang}}]{Fan_2020_constructing}%
  \BibitemOpen
  \bibfield  {author} {\bibinfo {author} {\bibfnamefont {R.-H.}\ \bibnamefont {Fan}}, \bibinfo {author} {\bibfnamefont {B.}~\bibnamefont {Xiong}}, \bibinfo {author} {\bibfnamefont {R.-W.}\ \bibnamefont {Peng}},\ and\ \bibinfo {author} {\bibfnamefont {M.}~\bibnamefont {Wang}},\ }\bibfield  {title} {\bibinfo {title} {Constructing metastructures with broadband electromagnetic functionality},\ }\href {https://doi.org/10.1002/adma.201904646} {\bibfield  {journal} {\bibinfo  {journal} {Adv. Mater.}\ }\textbf {\bibinfo {volume} {32}},\ \bibinfo {pages} {1904646} (\bibinfo {year} {2020})}\BibitemShut {NoStop}%
\bibitem [{\citenamefont {Kong}(1990)}]{Kong_1990_emwtheory}%
  \BibitemOpen
  \bibfield  {author} {\bibinfo {author} {\bibfnamefont {J.~A.}\ \bibnamefont {Kong}},\ }\href@noop {} {\emph {\bibinfo {title} {Electromagnetic {W}ave {T}heory}}}\ (\bibinfo  {publisher} {Wiley-Interscience},\ \bibinfo {year} {1990})\BibitemShut {NoStop}%
\bibitem [{\citenamefont {Lurie}\ \emph {et~al.}(2009)\citenamefont {Lurie}, \citenamefont {Onofrei},\ and\ \citenamefont {Weekes}}]{Lurie_2009_mathematical}%
  \BibitemOpen
  \bibfield  {author} {\bibinfo {author} {\bibfnamefont {K.}~\bibnamefont {Lurie}}, \bibinfo {author} {\bibfnamefont {D.}~\bibnamefont {Onofrei}},\ and\ \bibinfo {author} {\bibfnamefont {S.}~\bibnamefont {Weekes}},\ }\bibfield  {title} {\bibinfo {title} {Mathematical analysis of the waves propagation through a rectangular material structure in space--time},\ }\href {https://doi.org/10.1016/j.jmaa.2009.01.031} {\bibfield  {journal} {\bibinfo  {journal} {J. Math. Anal. Appl.}\ }\textbf {\bibinfo {volume} {355}},\ \bibinfo {pages} {180} (\bibinfo {year} {2009})}\BibitemShut {NoStop}%
\bibitem [{\citenamefont {Sejdi{\'c}}\ \emph {et~al.}(2009)\citenamefont {Sejdi{\'c}}, \citenamefont {Djurovi{\'c}},\ and\ \citenamefont {Jiang}}]{Sejdic_2009_time}%
  \BibitemOpen
  \bibfield  {author} {\bibinfo {author} {\bibfnamefont {E.}~\bibnamefont {Sejdi{\'c}}}, \bibinfo {author} {\bibfnamefont {I.}~\bibnamefont {Djurovi{\'c}}},\ and\ \bibinfo {author} {\bibfnamefont {J.}~\bibnamefont {Jiang}},\ }\bibfield  {title} {\bibinfo {title} {Time--frequency feature representation using energy concentration: An overview of recent advances},\ }\href {https://doi.org/10.1016/j.dsp.2007.12.004} {\bibfield  {journal} {\bibinfo  {journal} {Digit. Signal Process.}\ }\textbf {\bibinfo {volume} {19}},\ \bibinfo {pages} {153} (\bibinfo {year} {2009})}\BibitemShut {NoStop}%
\end{thebibliography}%
\clearpage

\onecolumngrid
\tableofcontents
\clearpage

\section{Pure-Time and Space-Time Discontinuities \\ in the Pulse and Continuous-Wave Regimes}\label{app:size_problem}
\begin{figure}[ht!]
    \centering
    \includegraphics[width=8.6cm]{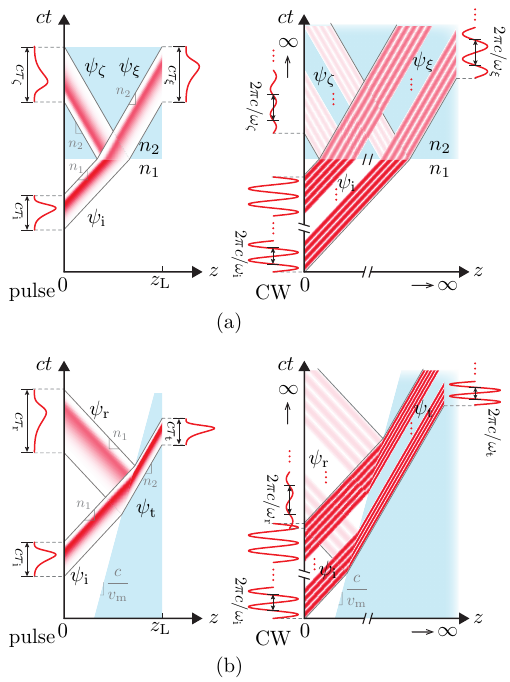}
    \caption{\label{fig:Pulse_CW} Scattering at a (a)~pure-time step interface and (b)~subluminal space-time step interface in the pulse-wave regime (left panels) and Continuous Wave (CW) regime (right panels).}
    \end{figure}

Figures~\ref{fig:Pulse_CW}(a) and (b) depict the wave scattering at a pure-time modulation step and space-time modulation step in both the pulse-wave regime (left panels) and continuous-wave regime (right panels). In the pulse-wave regime, a finite implementation size is possible, but synchronization is required between the incident wave and the modulation for scattering to occur. In the continuous-wave regime, the size of the device would need to extend to infinity to accommodate the infinite duration of the wave, as shown in the right panels of the figure.
    %

\clearpage
\section{Comparison of the Different Space-Time Structures}\label{app:comparision}
\begin{figure*}[ht!]
    \centering
    \includegraphics[width=17.8cm]{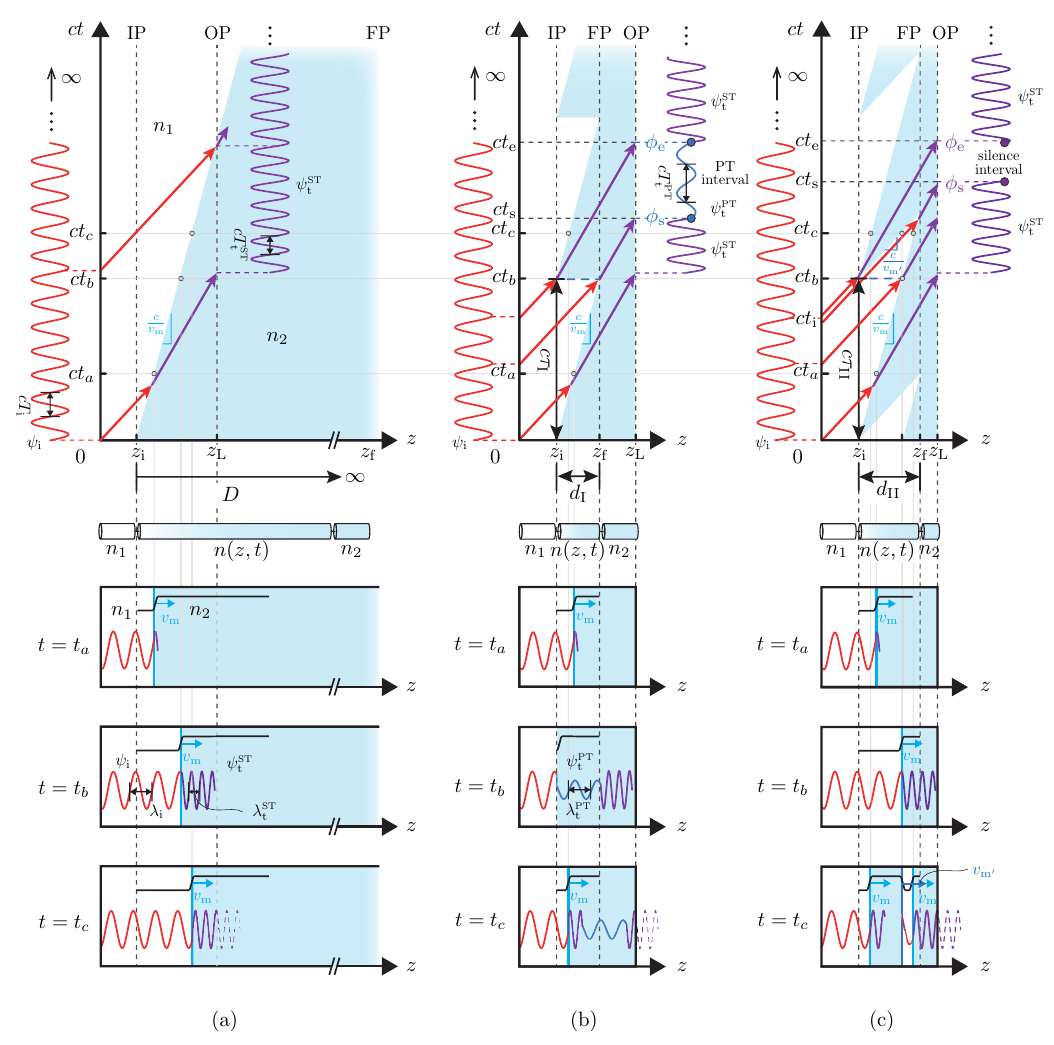}
    \caption{\label{fig:FP_MFP} Detailed operation principle of a (a) space-time interface [Fig.~\ref{fig:ss_st}(b), left panel], (b) Prism~I [Fig.~\ref{fig:Two_Types}(a)] and (c) Prism~II [Fig.~\ref{fig:Two_Types}(b)] in space-time (top panels) and space-index (middle and bottom panels) dimensions. IP: Initial Position; FP: Final Position; OP: Observation Position.}
    \end{figure*}

Figure~\ref{fig:FP_MFP} provides details on the operation principle of a space-time interface and the two proposed space-time prisms.

The top panels of Fig.~\ref{fig:FP_MFP} show 1+1~D space-time diagrams illustrating the transmission behavior across an infinite space-time interface compared with transmission across the two proposed types of space-time Fresnel prisms. Note that in the case of the infinite interface, represented in Fig.~\ref{fig:FP_MFP}(a), the load must be positioned before the end of the modulation, $z_{\mathrm{L}}<z_{\mathrm{f}}$ since $z_{\mathrm{f}}$ tends to infinity. In contrast, in the case of Prisms~I and~II, which are respectively represented in Figs.~\ref{fig:FP_MFP}(b) and (c), the load may be positioned at the end of the (now finite) structure, $z_{\mathrm{L}}>z_{\mathrm{f}}$.

The middle panels of Fig.~\ref{fig:FP_MFP} present a 1D optical waveguide (e.g., optical fiber) analogy, with source, modulation and load, and with the refractive indices varying along the axis of the waveguide.

The bottom panels of Fig.~\ref{fig:FP_MFP} represent the amplitude of the modulation and waves at three specific times, $t_{a}$, $t_{b}$ and $t_{c}$, corresponding to the space-time diagram in the top panels of the figure. The illustration provides insight into the modulation process and demonstrates how the wave amplitudes evolve over time and space. By examining the wave behavior at such different instants, one can gain a deeper understanding of the space-time characteristics and dynamics associated with the modulation.

\clearpage 
\section{Conversion Efficiency for the Two Prisms without Interconnections}\label{app:efficiency}
\begin{figure}[ht!]
    \centering
    \includegraphics[width=8.6cm]{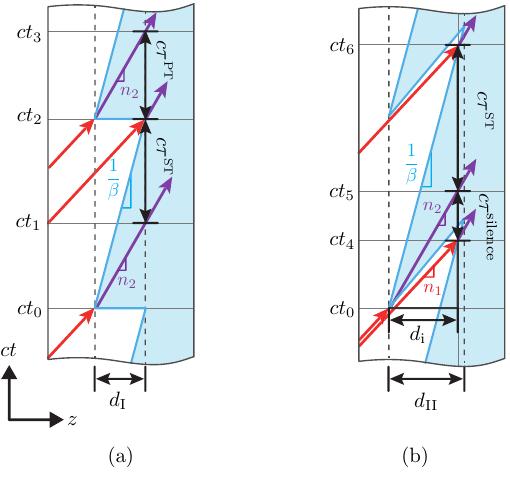}
    \caption{\label{fig:efficiency}Space-time diagrams to derive the conversion efficiency formulas for (a)~Prism~I and (b)~Prism~II.}
    \end{figure}

Figure~\ref{fig:efficiency} shows space-time diagrams of Prism~I and Prism~II with relevant information to derive their conversion efficiency formulas.

To derive the conversion efficiency for the first type of space-time Fresnel prism, depicted in Fig.~\ref{fig:efficiency}(a), we compute the Pure-Time (PT) interval $\tau^{\mathrm{PT}}$ and the Space-Time (ST) interval $\tau^{\mathrm{ST}}$, using space-time geometrical relations, which yields
    \begin{subequations}\label{aeq:tau_st_pt}
        \begin{equation}
            \tau^{\mathrm{PT}}=t_3-t_2=\frac{d_{\mathrm{I}}}{c} n_2
        \end{equation}
        and
        \begin{equation}
            \tau^{\mathrm{ST}}=(t_2-t_0)-(t_1-t_0)=\frac{d_{\mathrm{I}}}{c}(1/\beta- n_2),
        \end{equation}
        respectively.
    \end{subequations}
The conversion efficiency $\eta_{\mathrm{I}}$, is then found as
    \begin{equation}\label{aeq:eff_i}
        \eta_{\mathrm{I}}=\frac{\tau^\mathrm{ST}}{\tau^\mathrm{PT}+\tau^\mathrm{ST}}=\frac{1/\beta- n_2}{1/\beta}=1-n_2 \beta,
    \end{equation}
 where the second identity is obtained by substituting Eq.~\eqref{aeq:tau_st_pt} into Eq.~\eqref{aeq:eff_i}.

To derive the conversion efficiency for the second type of space-time Fresnel prism, shown in Fig.~\ref{fig:efficiency}(b), we compute the silence interval $\tau^{\mathrm{silence}}$ and the space-time interval $\tau^{\mathrm{ST}}$, using again space-time geometrical relations, which yields
    \begin{subequations}\label{aeq:tau_st_silence}
        \begin{equation}
            \tau^{\mathrm{silence}}=(t_5-t_0)-(t_4-t_0)=\frac{d_{\mathrm{i}}}{c} (n_2-n_1)
        \end{equation}
        and
        \begin{equation}
            \tau^{\mathrm{ST}}=(t_6-t_0)-(t_5-t_0)=\frac{d_{\mathrm{i}}}{c}(1/\beta- n_2),
        \end{equation}
        respectively, where $d_{\mathrm{i}}$ represents the catch-up distance between the trajectory of the part of the incident wave interacting with the left vertex of Prism~II and the moving interface with velocity $v_{\mathrm{m}}$.
    \end{subequations}
The conversion efficiency $\eta_{\mathrm{II}}$, is then found as
    \begin{equation}\label{aeq:eff_ii}
        \eta_{\mathrm{II}}=\frac{\tau^\mathrm{ST}}{\tau^\mathrm{silence}+\tau^\mathrm{ST}}=\frac{1/\beta- n_2}{1/\beta- n_1}=\frac{1- n_2 \beta}{1- n_1 \beta},
    \end{equation}
 where the second identity is obtained by substituting Eq.~\eqref{aeq:tau_st_silence} into Eq.~\eqref{aeq:eff_ii}.

\clearpage
\section{Comparison of Spurious Effects in the Conventional Fresnel Prism \\ and in the Space-Time Fresnel Prism}\label{app:spurious}

\begin{figure}[ht!]
    \centering
    \includegraphics[width=8.6cm]{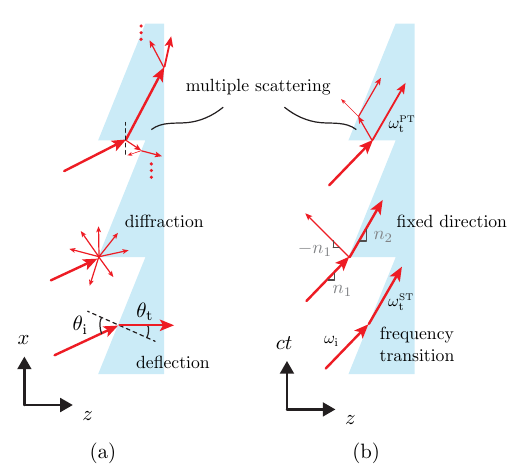}
    \caption{\label{fig:spurious} Desired (bottom sub-prisms) and spurious (middle and top sub-prisms) effects in (a)~the conventional Fresnel prism [Fig.~\ref{fig:ss_st}(a), right panel] and (b)~the space-time Fresnel prism with pure-time sections [Fig.~\ref{fig:Two_Types}(a)].}
    \end{figure}

Figure~\ref{fig:spurious} shows the space-space and space-time diagram for the desired and spurious effects in the conventional Fresnel prism [right panel of Fig.~\ref{fig:ss_st}(a)] and the space-time Fresnel prism with pure-time sections [Fig.~\ref{fig:Two_Types}(a)], respectively.

The arrows in the bottom sub-prisms show the desired deflection at a space-space interface [Fig.~\ref{fig:spurious}(a)] and frequency transition at a space-time interface [Fig.~\ref{fig:spurious}(b)], with reflections being neglected in both cases. 

The arrows in the middle sub-prisms represent the diffraction occurring at the apexes of the conventional Fresnel prism [Fig.~\ref{fig:spurious}(a)] (some diffraction, not shown, also occurs at the vertices on the right of the apexes) and the fixed-directional propagation occurring at the apexes of the space-time Fresnel prism [Fig.~\ref{fig:spurious}(b)], with small reflections being also shown for the sake of completeness. 

The arrows in the top sub-prisms depict the spurious transmission and multiple scattering occurring at the horizontal interfaces of both the conventional Fresnel prism [Fig.~\ref{fig:spurious}(a)] and the space-time Fresnel prism [Fig.~\ref{fig:spurious}(b)]. Note that, different from the conventional case, the multiple scattering in the space-time Fresnel prism is restricted to the positive vertical ($+ct$) direction due to causality.

\clearpage
\section{Phase-Matching Formulas for the Two Prisms}\label{app:FP}
\begin{figure}[ht!]
    \centering
    \includegraphics[width=11.5cm]{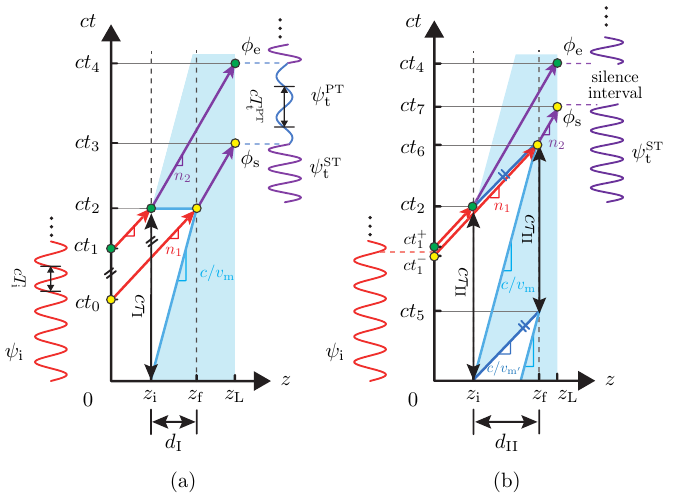}
    \caption{\label{afig:deri_FP_MFP} Space-time diagrams to derive the phase continuity design formulas for (a)~Prism~I and (b)~Prism~II.}
    \end{figure}

Figure~\ref{afig:deri_FP_MFP} shows space-time diagrams of Prism~I and Prism~II with relevant information to derive their phase matching design formulas. We selected here as the excitation a traveling-wave harmonic plane wave of electric field $E_{\mathrm{i}}=E_0\exp(-i\phi_{\mathrm{i}})$, with the phase $\phi_{\mathrm{i}}(z,ct)=\omega_{\mathrm{i}}t-k_{\mathrm{i}}z$, but the forthcoming derivations may be extended to (modulated or unmodulated) pulse excitations of different shapes.
\subsection{Prism~I [Eq.~\eqref{eq:FP_dimensions}]}
To derive the size and temporal periods of Prism~I [Fig.~\ref{afig:deri_FP_MFP}(a)] for phase continuity, we may utilize the phase tracing method~\cite{Kong_1990_emwtheory,Lurie_2009_mathematical}, tracking back the transmission phase from a given space plane to the incidence space plane ($z=0$), for simplifying the derivations. We find
    \begin{equation} \label{aeq:FP_phase_trace}
        \phi_{\mathrm{s}}(z_{\mathrm{L}},ct_3)=\phi(z_{\mathrm{f}},ct_2)=\phi(0,ct_0)
        \quad\textrm{and}\quad
        \phi_{\mathrm{e}}(z_{\mathrm{L}},ct_4)=\phi(z_{\mathrm{i}},ct_2)=\phi(0,ct_1).
    \end{equation}
Successively substituting the last equalities of Eqs.~\eqref{aeq:FP_phase_trace} into the phase continuity condition $\phi_{\mathrm{e}}-\phi_{\mathrm{s}}=2\pi q$ ($q\in\mathbb{N}$) and setting the related phase difference to the corresponding phase difference for the incident wave, $\phi_{\mathrm{i}}(z,ct)=\omega_{\mathrm{i}}t-k_{\mathrm{i}}z$, in the plane $z=0$, we obtain
\begin{subequations}
    \begin{equation}
        \phi_{\mathrm{e}}-\phi_{\mathrm{s}}=\phi(0,ct_1)-\phi(0,ct_0)=2\pi q=\omega_{\mathrm{i}}(t_1-t_0),
    \end{equation}
    so that
    \begin{equation}\label{aeq:i_phi_e_s}
        t_1-t_0=2\pi q/\omega_{\mathrm{i}}.
    \end{equation}
\end{subequations}
We obtain then the sought-after size and temporal period of Prism~I using space-time geometrical relations with Eq.~\eqref{aeq:i_phi_e_s}
    \begin{equation}\label{aeq:i_d_t}
        d_\mathrm{I}=z_{\mathrm{f}}-z_{\mathrm{i}}=(t_1-t_0)c/n_1=q\frac{\lambda_0}{n_1}
        \quad\text{and}\quad
        \tau_\mathrm{I}=\frac{d_\mathrm{I}}{v_{\mathrm{m}}}=q\frac{cT_0}{n_1v_{\mathrm{m}}},
    \end{equation}
where $\lambda_0=2\pi c/\omega_{\mathrm{i}}$ and $T_0=2\pi/\omega_{\mathrm{i}}$ are the free-space incident wavelength and period, respectively.
\subsection{Prism~II [Eq.~\eqref{eq:MFP_dimensions}]}
For Prism~II [Fig.~\ref{afig:deri_FP_MFP}(b)], we use again the phase tracing method but this time applied to the silence intervals, which yields
    \begin{subequations} \label{aeq:MFP_phase_trace}
        \begin{equation}
            \phi_{\mathrm{s}}(z_{\mathrm{L}},ct_7)=\phi(z_{\mathrm{f}},ct_6)=\phi(0,ct^{-}_1)
        \end{equation}
        and
        \begin{equation}
            \phi_{\mathrm{e}}(z_{\mathrm{L}},ct_4)=\phi(z_{\mathrm{i}},ct_2)=\phi(0,ct^{+}_1),
        \end{equation}
    \end{subequations}
where $t^{\mp}_1=t_1\mp\delta t$ with $\delta t\rightarrow0$. Equation~\eqref{aeq:MFP_phase_trace} shows that the starting and ending phases of the silence intervals are inherently continuous, so that the phase continuity condition $\phi_{\mathrm{e}}=\phi_{\mathrm{s}}$ is automatically satisfied in Prism~II.

For a given prism size $d_{\mathrm{II}}$, the temporal period is given by
    \begin{equation}\label{aeq:MFP_T_1}
        \tau_{\mathrm{II}}=t_2-0=(t_6-0)-(t_5-0)
        =\left(\frac{1}{v_{\mathrm{m}}}-\frac{1}{v_{\mathrm{m}'}}\right)d_{\mathrm{II}}.
    \end{equation}
%
\clearpage
\section{Design of the Interconnected Prisms}\label{app:delay_FP}

\begin{figure}[ht!]
    \centering
    \includegraphics[width=11.5cm]{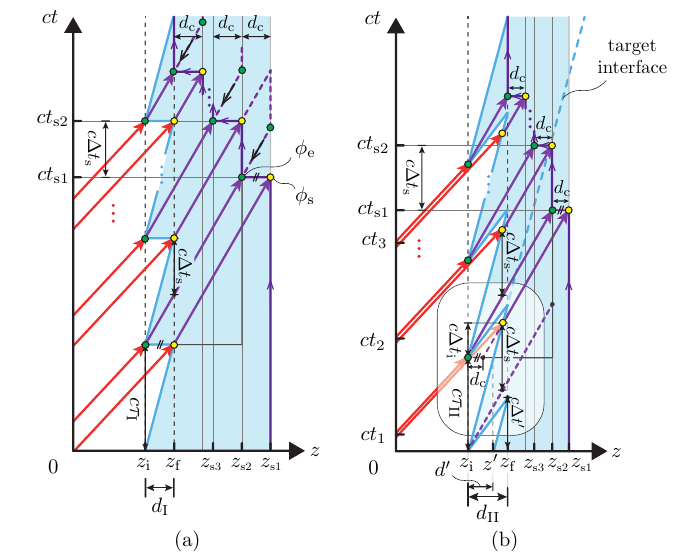}
    \caption{\label{afig:deri_delay_MFP} Space-time diagrams to derive the interconnection design formulas for (a)~Prism I and (b)~Prism~II with $N$ sub-prisms.}
    \end{figure}

Figure~\ref{afig:deri_delay_MFP} shows space-time diagrams of Prism~I and Prism~II with relevant information to derive their interconnection design formulas.
\subsection{Prism~I}
The design of interconnected Prism~I follows that of Prism~II in the body of the paper, based on space-time back-translating the points related to the transmitted phases $\phi_{\mathrm{e}}$ along the transmission trajectories to the same time levels as the $\phi_{\mathrm{s}}$-related points. We find, using the space-time geometrical relations in Fig.~\ref{afig:deri_delay_MFP}(a), that the interconnection distance $d_{\mathrm{c}}$ is equal to the prism size $d_{\mathrm{I}}$ [Eq.~\eqref{eq:FP_dimensions}], which yields
    \begin{equation}\label{aeq:FP_dc}
        d_{\mathrm{c}}=d_{\mathrm{I}}=q \frac{\lambda_0}{n_1}, \quad q\in\mathbb{N}.
    \end{equation}

Assuming that the first switch is positioned at $z_{\mathrm{s}1}$, the second switch must to be positioned to its left at a distance of $d_{\mathrm{c}}$ for phase continuity, i.e., at $z_{\mathrm{s}2} = z_{\mathrm{s}1}-d_{\mathrm{c}}$. The position of the third switch follows as $z_{\mathrm{s}3}=z_{\mathrm{s}2}-d_{\mathrm{c}}=z_{\mathrm{s}1}-2d_{\mathrm{c}}$, and so on, until the last switch ($N^{\mathrm{th}}$) reaches the final modulation position, $z_{\mathrm{s}N}=z_{\mathrm{f}}$. The $m^{\mathrm{th}}$ position of the switch is then given as
    \begin{equation}\label{eq:delay_FP_zlm}
        z_{\mathrm{s}m}=z_{\mathrm{f}}+(N-m) d_{\mathrm{c}},\quad m=1,2,3,...,N,
    \end{equation}
and the switching time of the $m^{\mathrm{th}}$ switch is hence
    \begin{subequations}
        \begin{equation}\label{aeq:FP_tsm}
            t_{\mathrm{s}m}=t_{\mathrm{s}1}+(m-1) \Delta t_{\mathrm{s}},
        \end{equation}
    with the switching time of the first switch being, according to Fig.~\ref{afig:deri_delay_MFP}(a),
        \begin{equation}
            t_{\mathrm{s}1}=\tau_{\mathrm{I}}+\frac{z_{\mathrm{s}2}-z_{\mathrm{i}}}{c/n_2}=\left(\frac{1}{v_{\mathrm{m}}}+\frac{N-1}{c/n_2}\right)d_{\mathrm{I}},
        \end{equation}
    where the second identity is obtained by substituting Eq.~\eqref{eq:delay_FP_zlm} with $m=2$ into $z_{\mathrm{s}2}$ and Eq.~\eqref{aeq:i_d_t} into $\tau_{\mathrm{I}}$ and rewriting the resulting equation in terms of $d_{\mathrm{I}}$,
    and where the (uniform) switching interval between the switching times, according to Fig.~\ref{afig:deri_delay_MFP}(a), is
        \begin{equation}
            \Delta t_{\mathrm{s}}=\tau_{\mathrm{I}}-\frac{d_{\mathrm{I}}}{c/n_2}=\left(\frac{1}{v_{\mathrm{m}}}-\frac{1}{c/n_{\mathrm{2}}}\right)d_{\mathrm{I}}.
        \end{equation} 
    \end{subequations}

\subsection{Prism~II [Eqs.~\eqref{eq:delay_distance}-\eqref{eq:delay_tsm}]}  
The derivation of the interconnection parameters $z_{\mathrm{s}m}$ [Eq.~\eqref{eq:delay_distance}] and $t_{\mathrm{s}m}$ [Eq.~\eqref{eq:delay_tsm}] for interconnected Prism~II [Fig.~\ref{afig:deri_delay_MFP}(b)] is fairly complicated and requires the space-time geometrical derivation of some preliminary formulas, which are the duration of the interluminal interface,
    \begin{subequations}
        \begin{equation} \label{aeq:ii_ta}
            \Delta t'=\frac{1}{v_{\mathrm{m}'}}d_{\mathrm{II}},
        \end{equation}
    the distance from the apex of the first two interfaces to the third interface,
        \begin{equation}\label{aeq:MFP_da}
            d'=\Delta t'(v_{\mathrm{m}'}-v_{\mathrm{m}})=\frac{v_{\mathrm{m}'}-v_{\mathrm{m}}}{v_{\mathrm{m}'}}d_{\mathrm{II}},
        \end{equation}
    where the second relation is obtained by using Eq.~\eqref{aeq:ii_ta}, and the delay of the incident wave trajectory between the apex of the first two interfaces and the third interface,
        \begin{equation}\label{aeq:delay_MFP_dti}
            \Delta t_{\mathrm{i}}=\frac{d'}{c/n_1-v_{\mathrm{m}}}=\frac{v_{\mathrm{m}'}-v_{\mathrm{m}}}{v_{\mathrm{m}'}(c/n_1-v_{\mathrm{m}})}d_{\mathrm{II}},
        \end{equation}
    where the second relation is obtained by using the second equality in Eq.~\eqref{aeq:MFP_da}.
    \end{subequations}

The interconnection distance $d_{\mathrm{c}}$ is then obtained, by phase back-tracing, as
    \begin{equation}\label{aeq:ii_dc}
        d_{\mathrm{c}}=\Delta t_{\mathrm{i}}(c/n_1-c/n_2)=\frac{(c/n_1-c/n_2)(v_{\mathrm{m}'}-v_{\mathrm{m}})}{v_{\mathrm{m}'}(c/n_1-v_{\mathrm{m}})}d_{\mathrm{II}},
    \end{equation}
    where the second identity is obtained by substituting Eq.~\eqref{aeq:delay_MFP_dti} and rewriting the equation in terms of $d_{\mathrm{II}}$.

 The $m^{\mathrm{th}}$ position of the switch may next be expressed, similarly to the case of Prism~I, as
    \begin{equation}\label{eq:delay_ii_zsm}
        z_{\mathrm{s}m}=z_{\mathrm{f}}+(N-m) d_{\mathrm{c}},\quad m=1,2,3,...,N,
    \end{equation}
with the same switching time formula
    \begin{subequations}
        \begin{equation}\label{aeq:ii_tsm}
            t_{\mathrm{s}m}=t_{\mathrm{s}1}+(m-1) \Delta t_{\mathrm{s}},
        \end{equation}
    but different switching parameters
        \begin{equation}
            t_{\mathrm{s}1}=\tau_{\mathrm{II}}+\frac{z_{\mathrm{s}2}-z_{\mathrm{i}}}{c/n_2}=\tau_{\mathrm{II}}+\frac{ d_{\mathrm{II}}+(N-2)d_{\mathrm{c}} }{c/n_2},
        \end{equation}
    where the second identity is obtained by substituting Eq.~\eqref{eq:delay_ii_zsm} with $m=2$ into $z_{\mathrm{s}2}$ with $\tau_{\mathrm{II}}$ and $d_{\mathrm{c}}$ given in Eqs.~\eqref{aeq:MFP_T_1} and~\eqref{aeq:ii_dc}, respectively,
    and 
        \begin{equation}
            \Delta t_{\mathrm{s}}=\frac{c\Delta t_{\mathrm{i}}}{n_1} \left(\frac{1}{v_{\mathrm{m}}}-\frac{n_2}{c}\right)=\frac{n_2(v_{\mathrm{m}'}-v_{\mathrm{m}})(c/n_2-v_{\mathrm{m}})}{n_1 v_{\mathrm{m}} v_{\mathrm{m}'}(c/n_1-v_{\mathrm{m}})}d_{\mathrm{II}}.
        \end{equation}
    \end{subequations}
%

\clearpage
\section{Size Comparison of Interconnected Prism~II \\ and the Target Moving Interface}\label{app:delay_ratio}

We shall compare here the implementation size of interconnected Prism~II [Fig.~\ref{fig:Delay_MFP}] with that of the target moving interface [left panel of Fig.~\ref{fig:ss_st}(b)] for a finite number of sub-prisms, $N$, for the switched-delay-line scheme. 

Interconnected Prism~II [Fig.~\ref{afig:deri_delay_MFP}(b)], composed of $N$ sub-prisms and involving $N-1$ transmission line sections, has the extra delay-line size $(N-1)d_{\mathrm{c}}$ compared to the initial Prism~II size, $d_{\mathrm{II}}$ [Fig.~\ref{fig:Two_Types}(b)]. The total size of the interconnected Prism~II is then obtained as
    \begin{equation}\label{aeq:ii_total_size}
        D_{\mathrm{II}}=d_{\mathrm{II}}+(N-1)d_{\mathrm{c}},
    \end{equation}
with $d_{\mathrm{c}}$ given in Eq.~\eqref{aeq:ii_dc}.

Within the time duration corresponding to $N$ sub-prisms in interconnected Prism~II, the target moving interface with velocity $v_{\mathrm{m}}$ reaches the size [Fig.~\ref{afig:deri_delay_MFP}(b)]
    \begin{equation}\label{aeq:interface_D}
        D=v_{\mathrm{m}}~(N \tau_{\mathrm{II}}+\Delta t'),
    \end{equation}
with $\tau_{\mathrm{II}}$ and $\Delta t'$ given in Eqs.~\eqref{aeq:MFP_T_1} and~\eqref{aeq:ii_ta}. 

We may then compute the size reduction ratio as
    \begin{equation}\label{aeq:size_ratio_0}
        \gamma=\frac{D_{\mathrm{II}}}{D}=\frac{v_{\mathrm{m}'}(c/n_1-v_{\mathrm{m}})+(N-1)(c/n_1-c/n_2)(v_{\mathrm{m}'}-v_{\mathrm{m}})}{(c/n_1-v_{\mathrm{m}})[N v_{\mathrm{m}'}-(N-1)v_{\mathrm{m}}]},
    \end{equation}
where the second identity is obtained by substituting Eq.~\eqref{aeq:ii_dc} into Eq.~\eqref{aeq:ii_total_size} and Eqs.~\eqref{aeq:MFP_T_1} and~\eqref{aeq:ii_ta} into Eq.~\eqref{aeq:interface_D}, respectively.    

For simplicity, we set the velocity of the interluminal interface to $v_{\mathrm{m}'}=c/n_1$. Then Eq.~\eqref{aeq:size_ratio_0} reduces to
    \begin{equation}\label{aeq:size_ratio}
        \gamma=\frac{Nn_2/n_1-(N-1)}{Nn_2/n_1-(N-1)\beta n_2}.
    \end{equation}
In the subluminal regime, where $\beta n_2<1$, the numerator of Eq.~\eqref{aeq:size_ratio} is always smaller than the denominator, so that the size reduction ratio is always smaller than $1$. This implies that the implementation size of interconnected Prism~II is always smaller than the size of the target moving interface.

\begin{figure}[ht!]
    \centering
    \includegraphics[width=8.6cm]{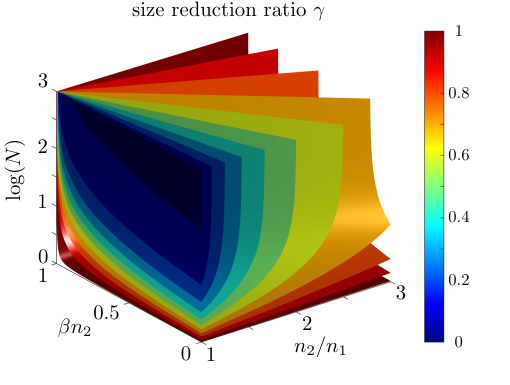}
    \caption{\label{afig:size_reduction} Isosurface plot for the size reduction ratio, $\gamma$ [Eq.~\eqref{aeq:size_ratio}], of interconnected Prism~II versus the modulation strength $n_2/n_1$, the normalized (to the wave velocity in the second medium) modulation velocity $\beta n_2$ and the logarithm of the number of sub-prisms. The interluminal interface velocity is chosen as $v_{\mathrm{m}'}=c/n_1$ for simplicity.}
    \end{figure}

To quantitatively visualise the reduction effect, Fig.~\ref{afig:size_reduction} plots the reduction ratio $\gamma$ versus the modulation strength $n_2/n_1$, the normalized (to the wave velocity in the second medium) modulation velocity $\beta n_2$ and the logarithm of the number of sub-prisms. A structure with a small reduction ratio can be realized by properly choosing the values of $n_1$, $n_2$ and $\beta$.
%

\clearpage
\section{Additional Simulation Results}\label{app:fdtd}
\begin{figure}[ht!]
    \centering
    \includegraphics[width=17.8cm]{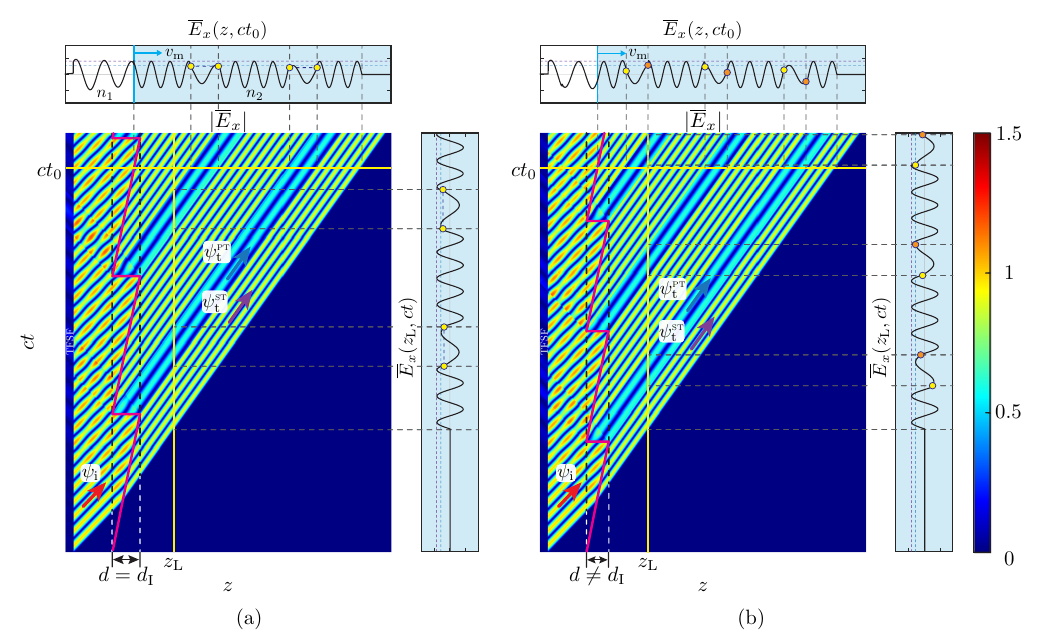}
    \caption{\label{fig:sup_FDTD_FP} Full-wave simulated (FDTD) normalized electric field ($|\overline{E}_{x}|=|E_{x}/E_0|$) for Prism~I (a)~with the phase continuity condition [Eq.~\eqref{eq:FP_dimensions}] satisfied and (b)~with the phase continuity condition unsatisfied, for the parameters $\epsilon_1=1.1$, $\epsilon_2=2$, $\mu=1$ and $v_{\mathrm{m}}=0.2c$. In both case, the insets show the field waveforms at a specific space position $z_{\mathrm{L}}$ (right inset) and time point $t_0$ (top inset).}
    \end{figure}

To validate the phase continuity formula for Prism~I, we present corresponding field plots for two scenarios in Fig.~\ref{fig:sup_FDTD_FP}: (i)~$d=d_{\mathrm{I}}$ [phase matching size, according to Eq.~\eqref{eq:FP_dimensions}], and (ii)~$d\neq d_{\mathrm{I}}$. 

\begin{figure}[ht!]
    \centering
    \includegraphics[width=8.6cm]{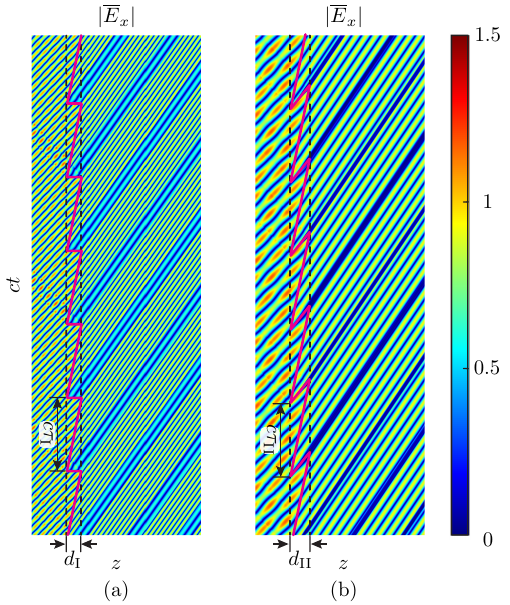}
    \caption{\label{fig:sup_FDTD_fft} Full-wave simulated (FDTD) normalized electric field ($|\overline{E}_{x}|=|E_{x}/E_0|$) for (a)~Prism~I [\ref{fig:Two_Types}(a)] and (b)~Prism~II [\ref{fig:Two_Types}(b)] with the parameters $\epsilon_1=1.1$, $\epsilon_2=2$, $\mu=1$, $v_{\mathrm{m}}=0.2c$ and $v_{\mathrm{m}'}=0.8 c$.}
    \end{figure}

Figure~\ref{fig:sup_FDTD_fft} shows the space-time diagrams for Prisms~I and II over a larger space-time area (more sub-prisms). For the sake of simplicity and clarity, the space-time diagrams are cropped to a finite space window.

\begin{figure}[ht!]
    \centering
    \includegraphics[width=8.6cm]{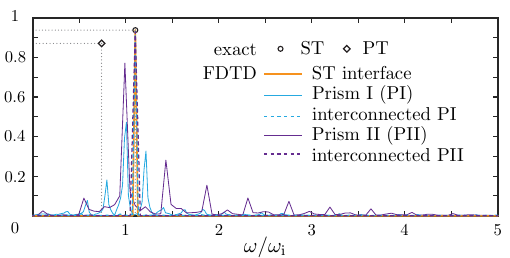}
    \caption{\label{fig:FDTD_sup} Full-wave simulated (FDTD) normalized electric field ($|\overline{E}_{x}|=|E_{x}/E_0|$) for the output waveform Fourier transforms for Prisms~I and~II in Fig.~\ref{fig:sup_FDTD_fft} without and with interconnections, and for the target space-time interface [Fig.~\ref{fig:ss_st}(b), left panel].}
    \end{figure}

Figure~\ref{fig:FDTD_sup} compares the output Fourier-transformed field distributions for the space-time Prisms without and with interconnections as well as for the target interface. Important differences -- including significant frequency and magnitude shifting as well as spurious sub-peaks -- occur for the case of the non-interconnected prisms, but this is because the related Fourier transforms were taken across the entire output waveforms, including the spurious pure-time and silence intervals, whereas the actual space-time waveform sections, which may be separately used in some pulse regime applications, naturally exhibit a much cleaner spectral response (obtainable by short-term Fourier transformation~\cite{Sejdic_2009_time}), as visualized in Figs.~\ref{fig:sup_FDTD_fft}(a) and (b). 

\end{document}